\documentclass[lettersize,journal,onecolumn]{IEEEtran}
 \pdfoutput=1
\usepackage{amsmath,amsfonts}
\usepackage{algorithmic}
\usepackage{algorithm}
\usepackage{array}
\usepackage[caption=false,font=normalsize,labelfont=sf,textfont=sf]{subfig}
\usepackage{textcomp}
\usepackage{stfloats}
\usepackage{url}
\usepackage{verbatim}
\usepackage{graphicx}
\usepackage{cite}
\usepackage{enumerate}
\usepackage{bm}
\usepackage{multirow}
\usepackage{color}
\usepackage{setspace}
\doublespace

\begin{document}

\title{Design of Joint Source-Channel Codes Based on a Generic Protograph \\
\thanks{
Part of the paper has been presented in ISTC 2021 \cite{lau2021joint}. \\
\indent The authors are with the Future Wireless Networks and IoT Focusing Area,
Department of Electronic and Information Engineering,
The Hong Kong Polytechnic University, Hong Kong SAR.
(Emails: jia1206.zhan@connect.polyu.hk and francis-cm.lau@polyu.edu.hk.) 
\\
\indent The work described in this paper was supported by a grant from the RGC
of the Hong Kong SAR, China (Project No.~PolyU 152170/18E).}
}

\author{\IEEEauthorblockN{Jia Zhan and Francis C.~M. Lau} }

%\IEEEauthorblockA{\textit{}
%\author{IEEE Publication Technology,~\IEEEmembership{Staff,~IEEE,}
%        % <-this % stops a space
%\thanks{This paper was produced by the IEEE Publication Technology Group. They are in Piscataway, NJ.}% <-this % stops a space
%\thanks{Manuscript received April 19, 2021; revised August 16, 2021.}}

% The paper headers
%\markboth{Journal of \LaTeX\ Class Files,~Vol.~14, No.~8, August~2021}%
%{Shell \MakeLowercase{\textit{et al.}}: A Sample Article Using IEEEtran.cls for IEEE Journals}
%
%\IEEEpubid{0000--0000/00\$00.00~\copyright~2021 IEEE}
% Remember, if you use this you must call \IEEEpubidadjcol in the second
% column for its text to clear the IEEEpubid mark.

\maketitle

\begin{abstract}
In this paper, we propose using a generic protograph to design joint source-channel codes (JSCCs). 
We present a generalized algorithm,
called protograph extrinsic information transfer for JSCC algorithm (PEXIT-JSCC algorithm),
 for analyzing the channel threshold of 
the proposed JSCC.
We also propose a source generic protograph EXIT (SGP-EXIT) algorithm, which is more appropriate than the extended source protograph extrinsic information transfer (ESP-EXIT) algorithm
for evaluating the source threshold of a generic protograph. 
Moreover, 
a collaborative optimization method based on the SGP-EXIT and PEXIT-JSCC algorithms is proposed to construct generic-protograph JSCCs with good source and channel thresholds.
Finally, we construct generic-protograph JSCCs, analyze their decoding thresholds, and compare their theoretical and error performance with JSCC systems based on optimized
double-protographs. Results show that our proposed codes can outperform 
double-protograph-based JSCCs.

%The generic protograph-based joint source-channel codes (JSCCs) can achieve comparable performance than optimized double protograph-based JSCCs by taking AR3A-JSCC and AR4JA-JSCC as examples. To design a generic protograph with both better waterfall and error floor performance compared with AR3A-JSCC and AR4JA-JSCC, we consider the channel threshold, the source threshold, and the linear minimum distance property at the same time.  After we obtain the generic protographs optimized, we can use the asymptotic weight distribution (AWD) tool to analyze the linear minimum distance properties of these codes. We generate some generic protographs based on the optimization method, and both their theoretical thresholds and error rate simulation results have shown they can achieve better error performance than AR3A-JSCC and AR4JA-JSCC and optimized double protographs.   
\end{abstract}


\section{Introduction}

{\color{black}In a digital communication system, source coding is employed to reduce the redundancy in the original information by compression while channel coding is used to protect the compressed data during transmission by adding redundant information (parity check bits). Traditionally, these two types of coding are 
studied and optimized separately. 
In particular, when the code length is very large (approaching infinity), the separate design of source and channel coding can theoretically achieve the optimal error performance over an additive white Gaussian noise (AWGN) channel according to the Shannon information theory \cite{shannon1948mathematical}. 
However, most real application scenarios cannot afford 
very long code lengths. 
They 
require low encoding/decoding latency and hence prefer short to moderate code lengths. 
With the recent development of Internet of Things, there has been a growing interest in combining source coding 
and channel coding with an aim to simplifying the system and/or further 
optimizing the transmission efficiency and effectiveness. 

The concept of joint source-channel coding (JSCC) was first conceived more than 40 years ago \cite{Robert1977}. It has been further investigated since 1990's \cite{b3,hagenauer1995source,b5}. The main idea is to allow the residual redundancy left by 
the source encoder to be utilized in the tandem joint source-channel decoding algorithms. 
For example, it is shown that 
considerable coding gains can be obtained
 by providing the prior probability of the source bits to the channel decoder \cite{hagenauer1995source}. 
Also when the redundancy of images is utilized in the decoding process, 
errors can be reduced \cite{burlina1998error}. 
In \cite{bi2017joint}, the use of JSCC in JPEG2000 transmission over a two-way relay network is studied; while
in \cite{Schaar2003, Martini2007}, the applications of JSCC schemes  to wireless video transmissions 
are investigated.
In \cite{Zribi2012}, an iterative joint source-channel decoder is proposed, where messages are exchanged between the decoder of variable length (source) codes (VLC) and the decoder of recursive systematic convolutional codes or low-density parity-check (LDPC) codes. With the use of this iterative joint source-channel decoder, significant 
error performance gains are observed compared with tandem decoding. 
 
% LDPC codes have the superior error-correction ability \cite{mackay1996near,richardson2001capacity,ten2004design}. Based on this point, researchers in \cite{fre2010joint} explored the feasibility of applying LDPC Codes to the JSCC system. And, a JSCC scheme, which adopted two low-density parity-check (LDPC) codes as both source and channel codes and used a joint decoder, was proposed. The joint decoding approach is to produce the external message exchanges between the channel decoder and the source decoder based on the joint protograph.

Low-density parity-check (LDPC) codes form a type of capacity-approaching 
channel codes and have been successfully deployed in many communication systems \cite{mackay1996near,richardson2001capacity,ten2004design}. 
In \cite{b9,fre2010joint}, two LDPC codes are concatenated in series to form a new type of joint
source-channel code. Moreover, an iterative decoding algorithm is presented to 
illustrate the extrinsic information exchange between the source-LDPC decoder and 
the channel-LDPC decoder.
%In \cite{he2012joint}, protograph LDPC (P-LDPC) codes 
% (a sub-type of LDPC codes having  linear-minimum-distance property and  fast encoding/decoding structure \cite{divsalar2005protograph, divsalar2009capacity,fang2015survey}) have been proposed to replace the two LDPC codes in the above JSCC, forming a JSCC scheme based on 
%double protograph LDPC (DP-LDPC) codes.
}

\begin{figure}[t!]
	\centering
	\includegraphics[width=3.5in]{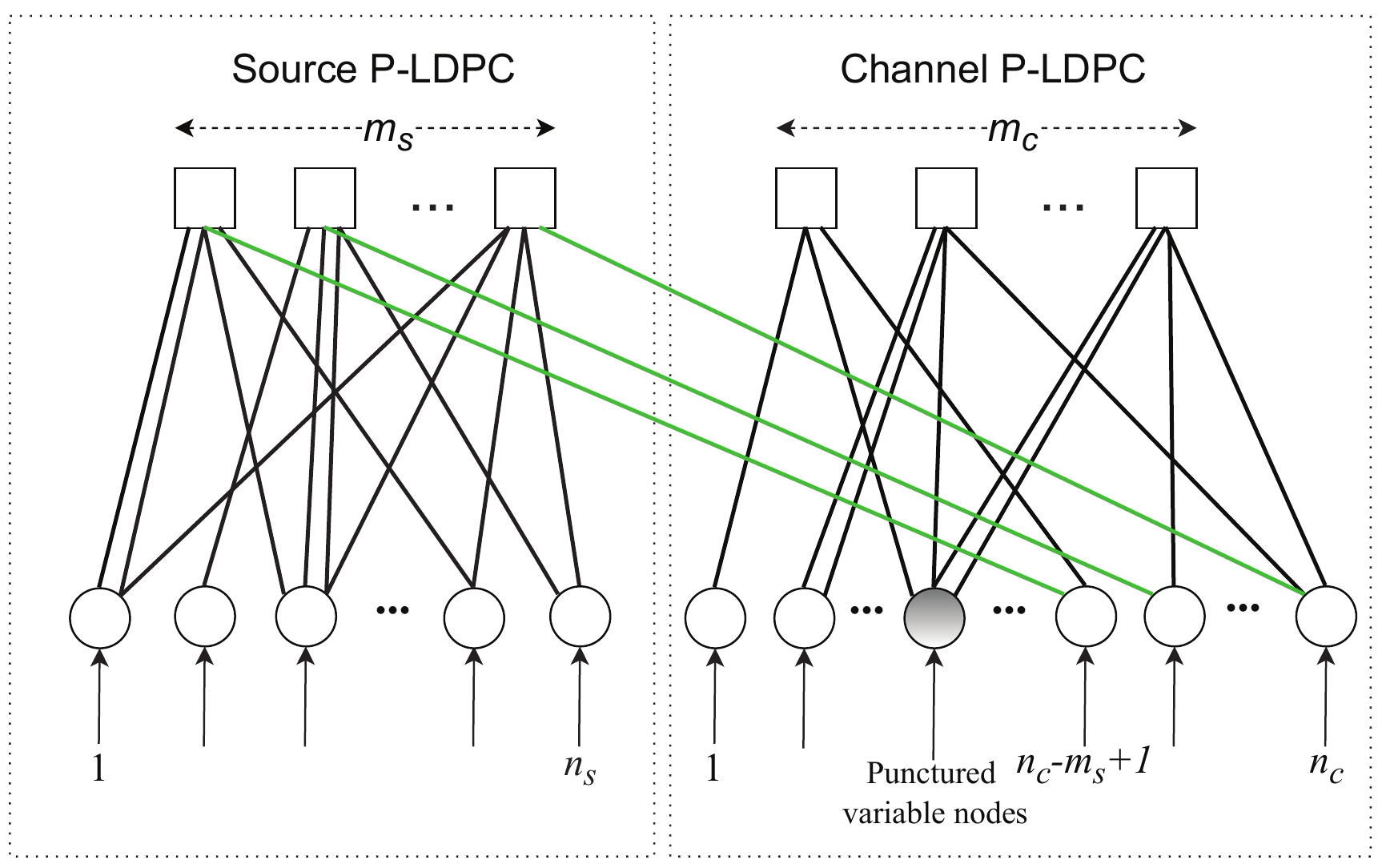}
	\caption{The protograph representation of a JSCC system in which two P-LDPC codes are cascaded. The source and channel protographs are depicted in the left and right dotted frames, respectively. Variable nodes and check nodes are, respectively, represented by circles and squares; and punctured variable nodes are represented by gray circles. The  green solid lines represent the cascading relationship between the source and channel encoders.
	}
	\label{JSCC_DPLDPC_origin}
\end{figure}

Protograph LDPC (P-LDPC) codes \cite{divsalar2005protograph,abbasfar2004accumulate,divsalar2009capacity,fang2015survey} are a subset of LDPC codes and they offer fast encoding and decoding structures as well as the linear minimum Hamming distance property. 
They have been proposed to replace the 
 regular LDPC codes in \cite{fre2010joint}, forming the double protograph-based LDPC JSCC 
 (DP-LDPC JSCC) system \cite{he2012joint}.
The protograph representation of this JSCC scheme is shown in Fig.~\ref{JSCC_DPLDPC_origin}. Variable nodes and check nodes are represented by circles and squares, respectively; and gray circles denote punctured variable nodes (VNs). The source and channel protographs are shown in the left and right dotted frames, respectively. Firstly, the source symbols are compressed based on the source (left)  protograph. Then, the compressed symbols are regarded as the inputs to the channel encoder. The solid green lines, which connect check nodes (CNs) in the source protograph and VNs in the channel protograph in a one-to-one manner, reflecting the cascading relationship between the source encoder and the channel encoder. Finally, the codeword is generated based on the channel (right) protograph. 

Fig.~\ref{JSCC_DPLDPC_origin} can be represented 
by a joint protomatrix $\textbf{B}_{J_0}$, i.e.,  
\begin{equation}
	\textbf{B}_{J_0}=
	\begin{pmatrix} 
	\makebox[-0.6cm]{} \textbf{B}_s &  \textbf{I}_{m_s} \ \textbf{0}_{m_s\times m_c} \\
	\;\; \textbf{0}_{m_c\times n_s} & \makebox[-0.5cm]{}  \textbf{B}_c
	 \end{pmatrix}
	 \label{eq:B_J0}
\end{equation}
where 
$\textbf{B}_s$ indicates the source protomatrix with size $m_s\times n_s$, 
$\textbf{B}_c$ indicates the channel protomatrix with size $m_c\times n_c$, 
$\textbf{I}_{m_s}$ is an identity matrix of size $m_s\times m_s$,
and $\textbf{0}$ denotes a zero matrix with size indicated by its subscript.
Optimizations on the DP-LDPC JSCC system has been performed under 
different scenarios. 
 In \cite{Wang2014} and \cite{He2017}, respectively,
an unequal error protection (UEP) technique 
and an unequal power allocation scheme have been proposed and applied 
to the DP-LDPC JSCC system. 
In \cite{chen2018design} and \cite{chen2016performance},  the source 
protomatrix $\textbf{B}_s$ and the channel protomatrix $\textbf{B}_c$, respectively,  have been redesigned in the JSCC system to improve the error performance. 
In \cite{chen2018joint}, 
 both source and channel protomatrices are redesigned at the same time to achieve good error-correction capability; while in \cite{Chen2019}, the optimal distribution of degree-2 VNs in both source and channel protomatrices is studied.
 In the above studies, error floors  are observed at high SNR for DP-LDPC JSCC systems 
 with relatively good performance in the waterfall region.

%In \cite{chen2016performance,chen2018joint,chen2018design}, 
%the source and channel P-LDPC codes in the JSCC system have been redesigned  to improve the error performance. 

\begin{figure}[t!]
	\centering
	\includegraphics[width=3.5in]{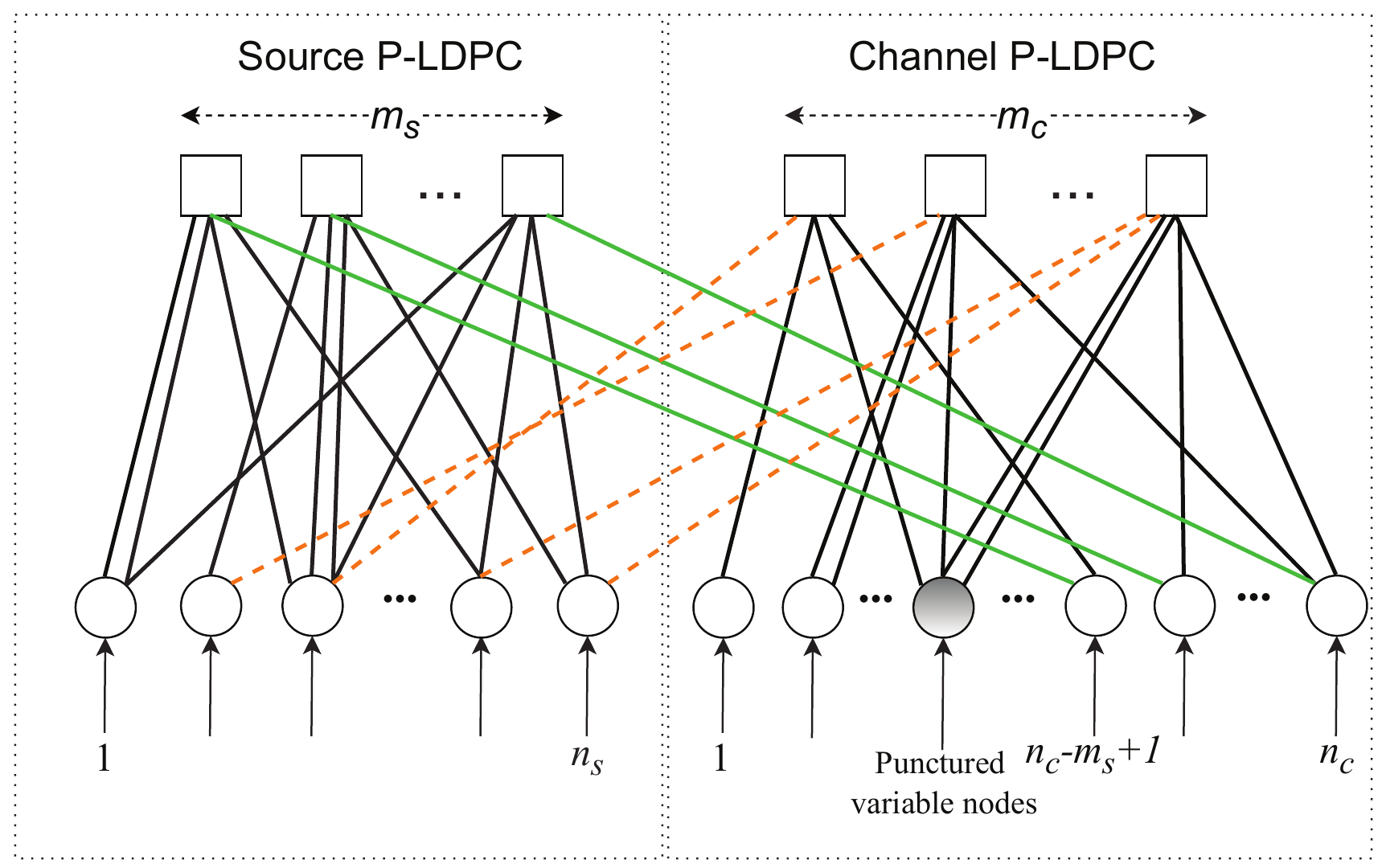}
	\caption{The protograph representation of the JSCC system where two P-LDPC codes are used and new edges (denoted by orange dashed lines) between variable nodes in the source protograph and check nodes in the channel protograph are added. 
	%The source and channel protographs are depicted in the left and right dotted frames, respectively. Variable nodes and check nodes are respectively represented by circles and squares, with gray circles representing punctured variable nodes.
}
	\label{JSCC_DPLDPC}
\end{figure}

In \cite{neto2013multi}, it is shown that the aforementioned error floors can be lowered 
by adding new edges between VNs in the source protograph and CNs in the channel protograph. At the same time, 
the error performance in the waterfall region will be moderately sacrificed. 
With the additional edges between VNs in the source protograph and CNs in the channel protograph, Fig.~\ref{JSCC_DPLDPC_origin} is modified to Fig.~\ref{JSCC_DPLDPC}
and
the joint protomatrix in \eqref{eq:B_J0} 
is modified to $\textbf{B}_{J}$, i.e.,
%In general, Fig.~\ref{JSCC_DPLDPC} can be used to illustrate the joint protograph in a DP-LDPC JSCC scheme. Moreover, it can be represented 
%by a joint protomatrix, i.e.,  
\begin{equation}
	\textbf{B}_{J}=\begin{pmatrix} 
	 \makebox[-0.4cm]{}  \textbf{B}_s & \textbf{B}_{sccv} \\
		 \textbf{B}_{svcc} &  \makebox[-0.4cm]{}  \textbf{B}_c \end{pmatrix}
	 \label{eq:B_J}
	 \end{equation}
where 
%$\textbf{B}_{sccv}= (\textbf{I}_{m_s} \ \textbf{0}_{m_s\times m_c})$
\begin{equation}
	\textbf{B}_{sccv}=\begin{pmatrix} \textbf{I}_{m_s} & \textbf{0} \end{pmatrix}
	\label{eq:constraint}
\end{equation}
 indicates the source-check-channel-variable linking protomatrix with size  $m_s\times n_c$
and $\textbf{B}_{svcc}$ indicates the source-variable-channel-check linking protomatrix with size $m_c\times n_s$  \cite{Hong2018,chen2020analysis}.
%Based on the joint protograph shown in Fig.~\ref{JSCC_DPLDPC}, researchers have also proposed many optimization methods. 
Subsequently,  the effect of $\textbf{B}_{svcc}$ on the code performance is studied.
In  \cite{Hong2018}, it has been discovered that 
connecting high-weight columns in $\textbf{B}_s$ to the rows
in $\textbf{B}_c$ via $\textbf{B}_{svcc}$ can obtain better error performance.
Moreover, if the identity matrix $\textbf{I}_{m_s}$ in $\textbf{B}_{sccv}$
is aligned with the 
 high-weight columns in $\textbf{B}_c$, 
 the error performance can be improved. 
 In other words, CNs in $\textbf{B}_s$ should be connected to VNs with high degrees in $\textbf{B}_c$ via $\textbf{I}_{m_s}$. 
 In \cite{QChen2019}, several design principles for optimizing $\textbf{B}_{svcc}$ are proposed to improve the waterfall performance when the source entropy is relatively high. 
In \cite{SLiu2019}, a search algorithm is proposed to find the
 best column permutation of $\textbf{B}_{sccv}$ given  $\textbf{B}_s$ and 
 $\textbf{B}_c$ are fixed. 
 (Note that permuting the columns of $\textbf{B}_{sccv}$ 
 while keeping $\textbf{B}_c$ fixed 
 is equivalent to permuting the columns of $\textbf{B}_c$ while keeping $\textbf{B}_{sccv}$ fixed.) 
In \cite{liu2020joint}, a joint optimization algorithm is provided to construct a joint protograph by taking the error floor and waterfall performance into account at the same time. 

Some theoretical analysis tools for the DP-LDPC JSCC system have also been proposed. A joint protograph extrinsic information transfer (JPEXIT) algorithm is proposed in \cite{chen2016performance} for calculating the channel threshold of the joint protograph shown in Fig.~\ref{JSCC_DPLDPC}. Lowering the channel threshold improves the waterfall performance. To evaluate the error-floor level, we also need to calculate the source threshold of the double protographs. In \cite{chen2015matching}, a source protograph extrinsic information transfer (SPEXIT) algorithm is proposed to calculate the source thresholds of DP-LDPC codes in the JSCC system when $\textbf{B}_{svcc}$ is a zero matrix. When $\textbf{B}_{svcc}$ is a non-zero matrix, a extended source protograph EXIT (ESP-EXIT) algorithm is proposed in \cite{chen2020analysis} to calculate the source thresholds of DP-LDPC codes.

As can be observed, all the aforementioned JSCC systems with DP-LDPC codes 
 have a structural constraint on $\textbf{B}_{sccv}$, which is shown in  \eqref{eq:constraint}. 
In this paper, we remove the above constraint. In other words, 
$\textbf{B}_{svcc}$ can be made up of arbitrary non-zero and zero entries. 
As a result, we view the 
so-called ``joint protomatrix $\textbf{B}_J$'' as 
a generic protomatrix, and name the corresponding JSCC
as ``generic protograph-based JSCC (GP-JSCC)'' \cite{lau2021joint}.
We consider the performance of the GP-JSCC in both the waterfall region and high signal-to-noise-ratio (SNR) region.
Since a lower channel threshold implies a better performance in the waterfall region,
we present a protograph EXIT for JSCC (PEXIT-JSCC) algorithm 
 for calculating the channel threshold of a generic P-LDPC (GP-LDPC) code in the JSCC system. 
It is also known that the source threshold affects the error floor in the high-SNR region. 
We therefore propose a source generic protograph EXIT (SGP-EXIT) algorithm, which is more generic than the ESP-EXIT algorithm  \cite{chen2020analysis},  for calculating the source threshold of a GP-LDPC code in the JSCC system.
In addition to the source threshold, the linear minimum distance property of a generic protograph can affect the error floor. 
Given a GP-LDPC code, we will apply the asymptotic weight distribution (AWD) tool \cite{divsalar2009capacity,divsalar2006ensemble,divsalar2005protograph} to calculate its typical minimum distance ratio (TMDR). 
If the code has a TMDR, there is a high probability that it possesses the linear minimum distance property, i.e., the minimum distance increases linearly with the codeword length. 
We illustrate our P-JSCC system by using 
`accumulate-repeat-4-jagged-accumulate" (AR4JA) and ``accumulate-repeat-3-and-accumulate" (AR3A) codes as examples  and forming AR4JA-JSCC and AR3A-JSCC codes  \cite{lau2021joint}. 

To design a P-JSCC with both good waterfall and error floor performance, 
we propose a joint optimization method which aims to achieve targeted source threshold and channel threshold.
Using AR4JA-JSCC and AR3A-JSCC as benchmarks, we further search for
  generic protographs in the JSCC schemes using the proposed joint optimization method. 
Finally, we compare the theoretical (source and channel) thresholds and simulated error performance of the 
generic protographs found 
with those of AR4JA-JSCC and AR3A-JSCC.
We also compare the results with those from optimized JSCC based on 
%Also, we have designed some generic protographs according to the proposed optimization method, and both theoretical thresholds and simulation results have shown they outperform AR3A-JSCC and AR4JA-JSCC and 
double protographs in \cite{liu2020joint}.

%\cite{lau2021joint}, such a constraint  has been  removed 
%and  a JSCC system based on a generic protograph is proposed.
%Furthermore, AR4JA-JSCC and AR3A-JSCC codes are formed by 
%using `accumulate-repeat-4-jagged-accumulate" (AR4JA) and ``accumulate-repeat-3-and-accumulate" (AR3A) codes as the generic protographs in the JSCC schemes.
%When compared with optimized double protographs in \cite{liu2020joint}, the AR3A-JSCC and AR4JA-JSCC codes have obtained comparable error performance. In this paper, we aim to conduct an in-depth study of the generic protograph-based JSCC system.
 
The main contributions of the paper are as follows.

\begin{enumerate}
\item We propose a JSCC scheme based on a generic protograph, namely 
the ``protograph-based JSCC (P-JSCC)''.
\item We present a generalized algorithm,
namely protograph EXIT for JSCC (PEXIT-JSCC) algorithm 
 for calculating the channel threshold of a GP-LDPC code in the JSCC system. 
 \item We propose a source generic protograph EXIT (SGP-EXIT) algorithm
for evaluating the source threshold of a generic protograph. 
The proposed technique can be used to calculate the source threshold of a double protograph, and the threshold value obtained is found to be the same as that calculated by the ESP-EXIT algorithm in \cite{chen2020analysis}. 
The ESP-EXIT algorithm, on the other hand, cannot be utilized to calculate the source threshold of a generic protograph in the JSCC system. Thus the proposed SGP-EXIT algorithm is more generic.
\item We propose a first-source-then-channel-thresholds (FSTCT) joint optimization method based on the SGP-EXIT and  PEXIT-JSCC algorithms. The objective is to obtain a generic protograph with a high source threshold and a low channel threshold at the same time. This method is implemented in two steps. 
{\color{black}The first step is to design 
a sub-protomatrix (related to the connections between the untransmitted VNs and the connected CNs) in a generic protomatrix to achieve a high source threshold based on the SGP-EXIT algorithm.} The second step is to design the remaining part of the generic protomatrix to achieve a low channel threshold based on the PEXIT-JSCC algorithm. Finally, we need to use the asymptotic weight distribution (AWD) tool to analyze the linear minimum distance property of the generic protomatrix/protograph.
\item {\color{black} By using the proposed joint optimization method, we construct some generic protomatrices
(P-JSCCs). 
Both theoretical analysis and error simulations reveal they can outperform AR3A-JSCC, AR4JA-JSCC and the optimized double protographs in \cite{liu2020joint}.}	
 \item Based on the simulation results, we have found that the waterfall performance becomes worse as the source entropy increases.
Moreover, a generic protograph designed for a given source entropy does not guarantee its optimality for other given source entropies. 
%Furthermore, when the given source entropy approaches a threshold, the error floor appears although no error floor occurs for relatively small source entropy.

%	\item{Theoretical thresholds and simulation results have also shown the waterfall performance becomes worse as the source information increases. The error floor of a generic protograph without the linear minimum distance property may or may not exist, and it is hard to be predicted. As the source information approaches a threshold value, the waterfall and error floor performance of a generic protograph becomes worse regardless of whether the code has the linear minimum distance property or not. Moreover, an optimized protograph for a given source entropy does not guarantee its optimality for other source entropies.}
 
\end{enumerate}

We organize this work as follows. 
Section~\ref{sect:P-JSCC} shows the details the proposed JSCC based on a generic protograph. 
In particular, we provide the structure of the P-JSCC, describe its encoding and decoding method, present a PEXIT-JSCC algorithm
 for analyzing its channel threshold, propose a SGP-EXIT algorithm
for evaluating its source threshold, and 
 propose a FSTCT joint optimization method for constructing good P-JSCCs. 
%Finally, we provide some conclusions and future work in Section~\ref{sect:concl}.}
%
%The  paper is organized as follows. 
%The system model of a JSCC system employing a GP-LDPC code is introduced in Section~\ref{two}. Section~\ref{three} provides an SGP-EXIT algorithm for calculating the source threshold of a generic protograph and introduces the joint optimization method. 
In Section~\ref{sect:results}, we present some optimized generic protographs and their theoretical thresholds. We further compare their error rate performance with JSCCs based on optimized double protographs with relevant discussions. Finally, we give some concluding remarks and some future directions in Section~\ref{sect:conclusion}.

\begin{figure}[!t]
	\centering
	\includegraphics[width=3.0in]{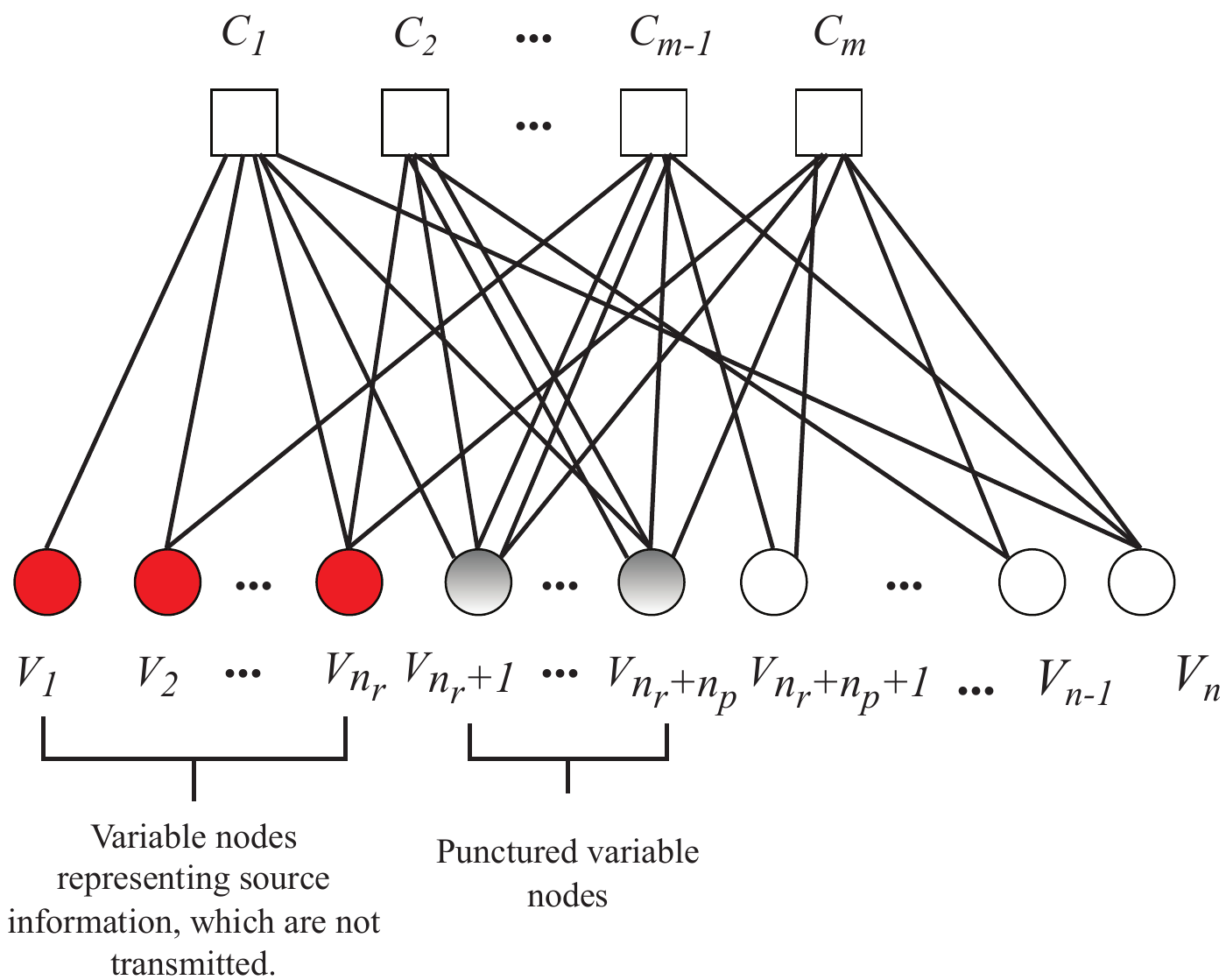}
	\caption{The proposed P-JSCC system using a generic protograph. 
	Variable nodes and check nodes are, respectively, represented by circles and squares, with gray circles indicating punctured variable nodes and red circles signifying variable nodes corresponding to source symbols.}
	\label{JSCC_SPLDPC}
\end{figure}

\section{Proposed JSCC system with a generic protograph} \label{sect:P-JSCC}
Referring to Fig.~\ref{JSCC_SPLDPC}, we illustrate the system model of the proposed P-JSCC system using a generic protograph that has no structural constraints. As shown in the figure,
 VNs and CNs are depicted by circles and squares, respectively. Punctured VNs are denoted by gray circles. VNs corresponding to source symbols are denoted by red circles. This generic protograph can alternatively be written as a generic protomatrix
 $\textbf{B}_{sp}$, i.e.,
\begin{equation}
\textbf{B}_{sp}=
\left(\begin{array}{*{20}{c}}
		{\begin{array}{*{20}{c}}
				{{e_{1,1}}} & {...} & {{e_{1,{n_r}}}} & {...} & {{e_{1,{n_r+n_p}}}} & {...} & {{e_{1,n}}}  \\
		\end{array}}  \\
		{\begin{array}{*{20}{c}}
				{{e_{2,1}}} & {...} & {{e_{2,{n_r}}}} & {...} & {{e_{2,{n_r+n_p}}}} & {...} & {{e_{2,n}}}  \\
		\end{array}}  \\
		\vdots   \\
		{\begin{array}{*{20}{c}}
				{{e_{m,1}}} & {...} & {{e_{m,{n_r}}}} & {...} & {{e_{m,{n_r+n_p}}}} & {...} & {{e_{m,n}}}  \\
		\end{array}}  \\
\end{array}\right)
\end{equation}
where $e_{i,j}$ denotes the $(i,j)$-th element ($i=1,2,..,m$, $j=1,2,...,n_r,...,n_r+n_p,...,n$). 
Here, 
$m$ denotes the total number of CNs; 
$n$ denotes the total number of VNs;
 $n_r$ denotes the number of VNs corresponding to source symbols;
  $n_p$ denotes the number of punctured VNs. 
% The structure in Fig.~\ref{JSCC_SPLDPC} is a generalization of the generic protograph-based JSCC model proposed in \cite{lau2021joint}.
The overall symbol code rate of the P-JSCC system is given by 
\begin{equation}
	\label{rate}
	R = n_r/(n-n_r-n_p).
\end{equation}

\begin{figure}[t]	
	\centering	
%	\subfigure[]
	{
		\label{Fig.1.a}
		\includegraphics[width=0.25\textwidth]{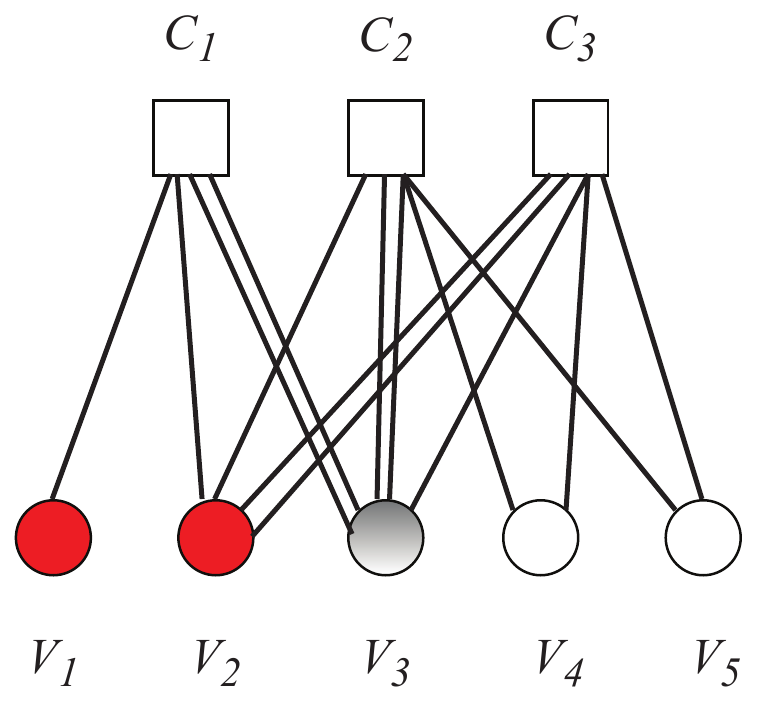}  \\ \vspace{1ex}
		(a) \\ \vspace{1ex}
		\centering
	}
%	\subfigure[]
	{
		\label{Fig.1.b}
		\includegraphics[width=0.25\textwidth]{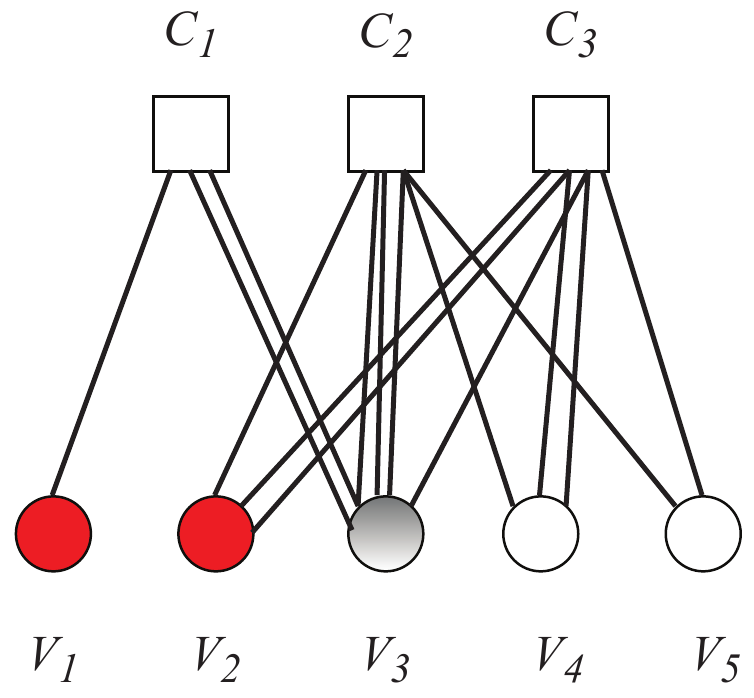}  \\ \vspace{1ex}
		(b) \\ %\vspace{1ex}
		\centering
	}	
	\caption{The protograph of (a) an AR3A-JSCC; (b) an AR4JA-JSCC. 
Variable nodes and check nodes are, respectively, represented by circles and squares, with gray circles indicating punctured variable nodes and red circles signifying variable nodes corresponding to source symbols.
%	
%	The first to fifth variable nodes are represented as $V_1, V_2, ..., V_5$, and the first to third check nodes are denoted by $C_1, C_2, C_3$. $V_1$ and $V_3$ denoted by solid red circles include the input source information and they are not transmitted to the receiver over channels. $V_2, V_4$ and $V_5$ contain parity information, and among them, $V_2$ represented by a blank circle is also punctured.
	}	
	\label{fig:codes}	
\end{figure}

\begin{figure}[t]	
	\centering	
%	\subfigure[]
	{
		\label{Fig.2.a}
		$\textbf{B}_{AR3A-JSCC}=\begin{array}{@{}r@{}c@{}c@{}c@{}c@{}c@{}l@{}}
			& V_1 & V_2 & V_3 & V_4 & V_5  \\
			\left.\begin{array}
				{c} C_1 \\C_2 \\C_3 \end{array}\right(
			& \begin{array}{c} 1 \\ 0 \\ 0 \end{array}
			& \begin{array}{c} 1 \\ 1 \\ 2 \end{array}
			& \begin{array}{c} 2 \\ 2 \\ 1 \end{array}
			& \begin{array}{c} 0 \\ 1 \\ 1 \end{array}
			& \begin{array}{c} 0 \\ 1 \\ 1 \end{array}
			& \left)\begin{array}{c} \\ \\ \\ \end{array}\right.
		\end{array}$ \\ \vspace{1ex}
		(a) \\ \vspace{1ex}
		\centering
	}
%	\subfigure[]
	{
		\label{Fig.2.b}
		$\textbf{B}_{AR4JA-JSCC}=\begin{array}{@{}r@{}c@{}c@{}c@{}c@{}c@{}l@{}}
			& V_1 & V_2 & V_3 & V_4 & V_5  \\
			\left.\begin{array}
				{c} C_1 \\C_2 \\C_3 \end{array}\right(
			& \begin{array}{c} 1 \\ 0 \\ 0 \end{array}
			& \begin{array}{c} 0 \\ 1 \\ 2 \end{array}
			& \begin{array}{c} 2 \\ 3 \\ 1 \end{array}
			& \begin{array}{c} 0 \\ 1 \\ 2 \end{array}
			& \begin{array}{c} 0 \\ 1 \\ 1 \end{array}
			& \left)\begin{array}{c} \\ \\ \\ \end{array}\right.
		\end{array}$\\ \vspace{1ex}
		(b) \\
		\centering
	}	
	\caption{The protomatrix of (a) an AR3A-JSCC code; (b) an AR4JA-JSCC code.}	
	\label{fig:matrices}	
\end{figure}	
%\begin{figure}[t]	
%	\centering	
%%	\subfigure[]
%	{
%		\label{Fig.2.a}
%		$\textbf{B}_{AR3A-JSCC}=\begin{array}{@{}r@{}c@{}c@{}c@{}c@{}c@{}l@{}}
%			& V_1 & V_2 & V_3 & V_4 & V_5  \\
%			\left.\begin{array}
%				{c} C_1 \\C_2 \\C_3 \end{array}\right(
%			& \begin{array}{c} 1 \\ 0 \\ 0 \end{array}
%			& \begin{array}{c} 2 \\ 2 \\ 1 \end{array}
%			& \begin{array}{c} 1 \\ 1 \\ 2 \end{array}
%			& \begin{array}{c} 0 \\ 1 \\ 1 \end{array}
%			& \begin{array}{c} 0 \\ 1 \\ 1 \end{array}
%			& \left)\begin{array}{c} \\ \\ \\ \end{array}\right.
%		\end{array}$ \\ \vspace{1ex}
%		(a) \\ \vspace{1ex}
%		\centering
%	}
%%	\subfigure[]
%	{
%		\label{Fig.2.b}
%		$\textbf{B}_{AR4JA-JSCC}=\begin{array}{@{}r@{}c@{}c@{}c@{}c@{}c@{}l@{}}
%			& V_1 & V_2 & V_3 & V_4 & V_5  \\
%			\left.\begin{array}
%				{c} C_1 \\C_2 \\C_3 \end{array}\right(
%			& \begin{array}{c} 1 \\ 0 \\ 0 \end{array}
%			& \begin{array}{c} 2 \\ 3 \\ 1 \end{array}
%			& \begin{array}{c} 0 \\ 1 \\ 2 \end{array}
%			& \begin{array}{c} 0 \\ 1 \\ 2 \end{array}
%			& \begin{array}{c} 0 \\ 1 \\ 1 \end{array}
%			& \left)\begin{array}{c} \\ \\ \\ \end{array}\right.
%		\end{array}$\\ \vspace{1ex}
%		(b) \\
%		\centering
%	}	
%	\caption{The protomatrix of (a) an AR3A-JSCC code; (b) an AR4JA-JSCC code.}	
%	\label{fig:matrices}	
%\end{figure}

\noindent \underline{Example}: 
 We adopt the conventional  AR3A code as
our proposed P-JSCC for illustration. 
Referring to Fig.~\ref{fig:codes}(a), our proposed P-JSCC, namely AR3A-JSCC, consists of 
\begin{itemize}
\item $m=3$ CNs ($C_1$ to $C_3$);
\item $n=5$ VNs ($V_1$ to $V_5$);
\item $n_r=2$ VNs corresponding to source symbols (indicated by red filled circles $V_1$ and $V_2$;
\item  $n_p=1$ punctured VNs (indicated by the gray circle $V_3$);
\item and an overall symbol code rate $R=2/(5-2-1)=1$. 
%
%variable nodes for source inputs and are punctured before transmission (indicated by red filled circles $V_1$ and $V_3$);
%\item unpunctured variable nodes in joint source-channel coding (indicated by black circles $V_4$ and $V_5$);
%\item punctured variable nodes in joint source-channel coding (indicated by the hollow  circles $V_2$);
%\item check nodes indicated by the symbol $\oplus$ ($C_1$ to $C_3$).
\end{itemize}
The corresponding protomatrix of the AR3A-JSCC code is shown in Fig.~\ref{fig:matrices}(a) and its size is $3\times 5$. 
When the conventional  AR4JA code is used in
our proposed P-JSCC, Fig.~\ref{fig:codes}(b) and Fig.~\ref{fig:matrices}(b) 
show, respectively, the protograph of the AR4JA-JSCC and its corresponding protomatrix.

%
%
%
%(Note that when $H_p$ changes, the size of the protograph and/or the above settings  
%may need to be adjusted to provide satisfactory performance.) 
% The overall code rate of the AR3A-JSCC code is given by
%	\begin{equation}\label{Rn}
%		R_{AR3A-JSCC}=\frac{m}{n-n^{'}_p}
%\end{equation}
%where $n$, $m$ and $n^{'}_p$  denote, respectively,  the total number of VNs, the number of VNs for source inputs, and the total number of  punctured VNs in the protograph. Since $n$, $m$ and $n^{'}_p$ in Fig.~\ref{Fig.1.a} are $5$, $2$ and $3$, respectively,  $R_{AR3A-JSCC}$ equals $1$.

%The generic protograph shown in Fig.~\ref{JSCC_SPLDPC} can also be regarded as a joint protograph shown in Fig.~\ref{JSCC_DPLDPC} without the constraint \eqref{constrain}. Thus the approaches of encoding, decoding and calculating channel thresholds in the generic protograph-based JSCC system is same as that in the double protograph-based JSCC system \cite{chen2016performance,liu2020joint}. The joint protograph extrinsic information transfer (JPEXIT) algorithm proposed in \cite{chen2016performance}, which is used to calculate the channel thresholds of double protographs with the constraint \eqref{constrain}, was redefined as the protograph extrinsic information transfer for JSCC (PEXIT-JSCC) algorithm in \cite{lau2021joint}. In \cite{lau2021joint}, the PEXIT-JSCC algorithm can be used to calculate the channel thresholds of the generic protographs with no structural constraints. The detailed introduction of the encoder, decoder and PEXIT-JSCC algorithm is shown as follows. 

\subsection{Encoder}\label{sect:P-JSCC encoder}

To generate an overall parity-check matrix, $\textbf{B}_{sp}$ can be lifted by the progressive-edge-growth (PEG) algorithm  \cite{hu2005regular}  
which can maximize the girth (i.e., smallest cycle) of the resultant Tanner graph.
Assume a lifting factor of $z$, the size of the lifted parity-check matrix, denoted by $\textbf{H}_{sp}$,
equals $mz \times nz$.
We assume a binary independent and identically distributed (i.i.d.) Bernoulli source. 
Denoting the probability of ``1" in the source sequence by $p_1$, the probability of ``0" in the source equals $1-p_1$. Therefore, the source entropy is given by
\begin{equation}\label{entropy}
	H_p = -p_1\log_2 p_1-(1-p_1)\log_2 (1-p_1),
\end{equation}
where $p_1\neq 0.5$.
For example, when the source probability $p_1= 0.04$, the source entropy $H_p=0.242$ bit/symbol.

\begin{figure}[t]
	\centerline{\includegraphics[keepaspectratio, width=0.45\textwidth]{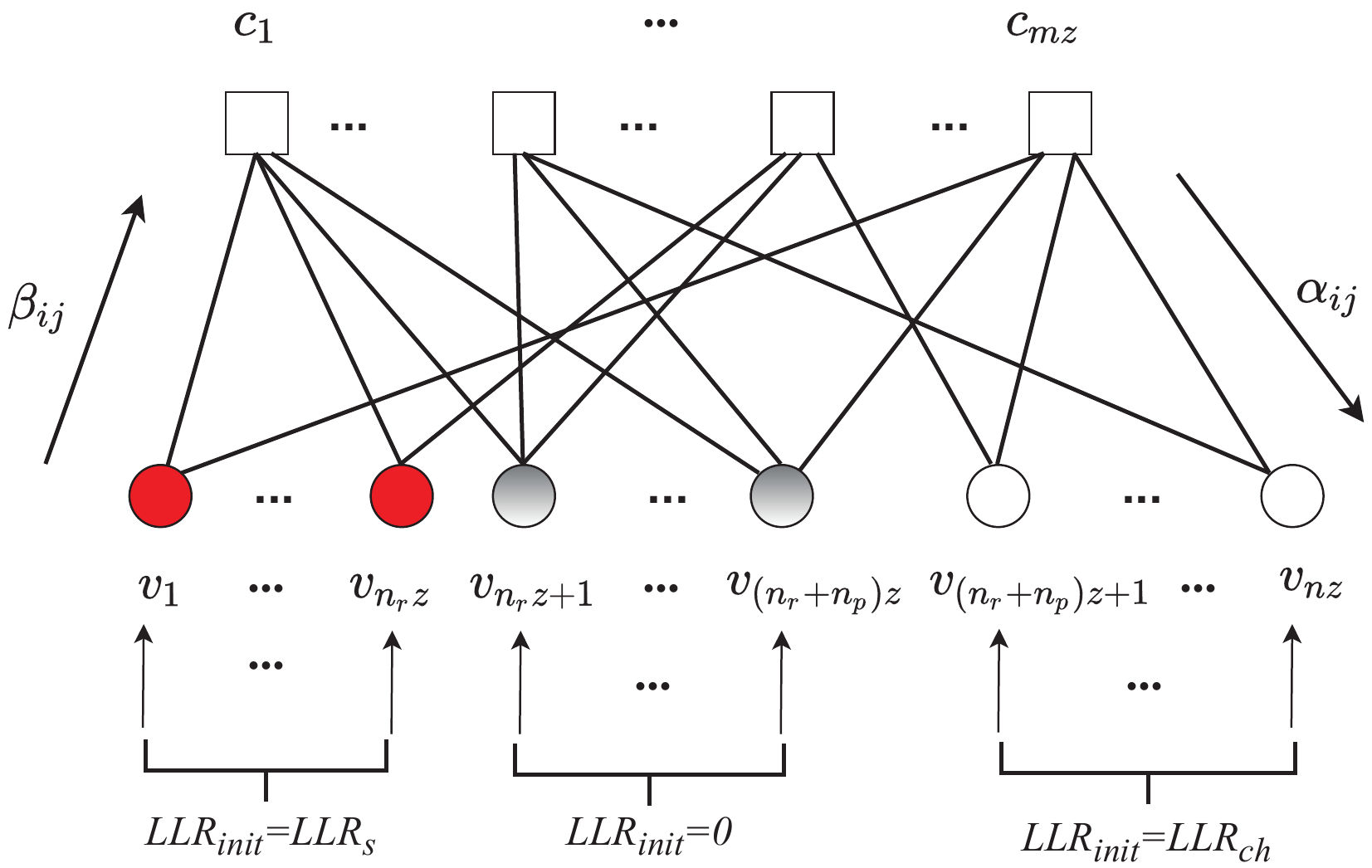}}
	\caption{Decoding of the P-JSCC.}
	\label{decoder}
\end{figure}

%To generate a large parity-check matrix $\textbf{H}_{sp}$ with a large girth, we employ the progressive-edge-growth (PEG) algorithm \cite{hu2005regular} to randomly lift the protomatrix $\textbf{B}_{sp}$. The lifting factor is denoted by $z$ and hence the size of $\textbf{H}_{sp}$ is $mz \times nz$. We assume that the source obeys the binary independent and identically Bernoulli distribution, and the probability with ``1'' in the source sequence is denoted by $p_1$ ($p_1 \neq 0.5$). Then, the source entropy is calculated by 
%
%\begin{equation}\label{entropy}
%	H = -p_1\log_2 p_1-(1-p_1)\log_2 (1-p_1).
%\end{equation}
To begin with, we generate a source sequence with a size of $1\times n_rz$ where the probability of ``1'' is $p_1$. The source is directly utilized as the input (referred to as $v_1$ to $v_{n_{r}z}$ VNs in Fig.~\ref{decoder}) to the joint encoder and generate the codeword $(v_1, \ldots, v_{nz})$ based on the parity-check matrix $\textbf{H}_{sp}$. Finally, the source symbols ($v_1$ to $v_{n_{r}z}$) and the punctured bits 
(i.e., $v_{n_{r}z+1}$ to $v_{(n_r+n_p)z}$ in Fig.~\ref{decoder}) are not transmitted while
the code bits $v_{(n_r+n_p)z+1}$ to $v_{nz}$ are sent.

\subsection{Decoder}\label{sect:P-JSCC decoder}
Fig.~\ref{decoder} illustrates the decoding of the P-JSCC. 
We define the following.
\begin{itemize}
	\item $I_{\max}$ is the maximum number of decoding iterations.
\item Binary phase-shift-keying modulation is used where bit ``$1$" and ``$0$" are mapped to ``$-1$" and ``$+1$", respectively. 
\item The noise variance of the AWGN channel is given by $\sigma^{2}$.
\item The received signal is given by $y=\pm1 + \eta$ where $\eta \sim N(0,\sigma^{2})$ denotes the AWGN.
\item $M$ and $N$, respectively, represent the number of CNs and VNs in the parity-check matrix $\textbf{H}_{sp}$. They equal $mz$ and $nz$, respectively, i.e., $M=mz$ and $N=nz$.
	\item $LLR_{init}(j)$ represents the initial log-likelihood-ratio (LLR) of the $j$-th VN ($j=1, 2, ..., N$). 
%	\item The index values of VNs and CNs are represented by $i$ ($i=1, 2, ..., M$) and $j$ ($j=1, 2, ..., N$) respectively.
	\item $\alpha_{ij}$ represents the LLR message sent from the $i$-th CN to the $j$-th VN.
	\item $\beta_{ij}$ denotes the LLR message sent from the $j$-th VN to the $i$-th CN. 
	\item $L_{APP, j}$ denotes the posterior LLR of the $j$-th VN.
	\item $\mathcal{M}(j)$ and $\mathcal{N}(i)$ represent the set of all CNs connected to the $j$-th VN and the set of all VNs connected to the $i$-th CN, respectively. 
	\item $\mathcal{M}(j) \backslash i$ denotes the set of all CNs connected to the $j$-th VN excluding the $i$-th CN; and $\mathcal{N}(i) \backslash j$ denotes the set of all VNs connected to the $i$-th CN excluding the $j$-th VN. 
\end{itemize}
The decoding process is described as follows. It is similar to that of decoding a traditional LDPC code except the
initialization process. 

\noindent {\textbf{Initialization}}:
\begin{itemize}
%	\item Set the iteration counter $r=0$.
	\item Set $\alpha_{ij}=\beta_{ij}=0$ $\forall i, j$.
	\item As shown in Fig.~\ref{decoder}, $LLR_{init}$ is given by
	\begin{equation}
		LLR_{init}(j) = \left\{
		\begin{array}{cc}
			LLR_{s}(j), & j=1,2,...,n_rz\\
		     0, & j=n_rz+1,n_rz+2,\\
		     ~ & ...,(n_r+n_p)z\\
		    LLR_{ch}(j), & j=(n_r+n_p)z+1,\\
		    ~ & (n_r+n_p)z+2,...,nz
		\end{array}
		\right.
	\end{equation}
    where $LLR_{s}(j)=\ln((1-p_1)/p_1)$ represents the initial LLR information of the source symbols; and $LLR_{ch}(j)=2 y_j/\sigma^{2}$ represents the initial LLR information from the channel.
\end{itemize}

\noindent {\textbf{Iterative process}}: 
\begin{itemize}
	\item Start: Set the iteration counter $r=1$. 
	\item Step 1) Updating LLRs from the VNs to the CNs: 
	\begin{equation}
		\beta_{ij} = LLR_{init}(j) + \sum_{i^{'}\in \mathcal{M}(j) \backslash i}
		\alpha_{i^{'}j}, \; \forall i,j
	\end{equation}
	\item  Step 2) Update LLRs from the CNs to the VNs 
	\begin{equation}
		\alpha_{ij} = 2\tanh^{-1}\Bigg(\prod_{j^{'}\in \mathcal{N}(i) \backslash j}\tanh(\beta_{ij^{'}}/2)\Bigg), \; \forall i, j.
	\end{equation}
	\item Step 3) Calculate the posterior LLRs by
	\begin{equation}
		L_{APP, j} = LLR_{init}(j) + \sum_{i\in \mathcal{M}(j)} \alpha_{ij}, \; \forall j.
	\end{equation}
	\item Step 4) Estimate $v_j$ by
	\begin{equation}
		{\hat{v}_j}=0 \text{ if } L_{APP,j} \geq 0,
		\text{ otherwise } {\hat{v}_j}=1, \; \forall j.  
	\end{equation}
	If	$\hat{\bm{v}}\cdot \textbf{H}_{sp}^{T}=\textbf{0}$ is satisfied
%	\begin{equation}\label{cond_J}
%		\hat{\bm{c}}\cdot \textbf{H}_{sp}^{T}=\textbf{0}
%	\end{equation}
	where $\hat{\bm{v}}=\{\hat{v}_1, \hat{v}_2, ..., \hat{v}_{nz}\}$,
% denotes the reconstructed codeword sequence. If Equation (\ref{cond_J}) is satisfied 
 or $r=I_{max}$, stop the iteration and output $\hat{\bm{v}}$ as the joint source-channel codeword; otherwise increase the iteration counter $r$ by 1, and repeat Step 1) to Step 4).
\end{itemize}

\subsection{PEXIT-JSCC algorithm}

\begin{figure}[t]
	\centerline{\includegraphics[keepaspectratio, width=0.45\textwidth]{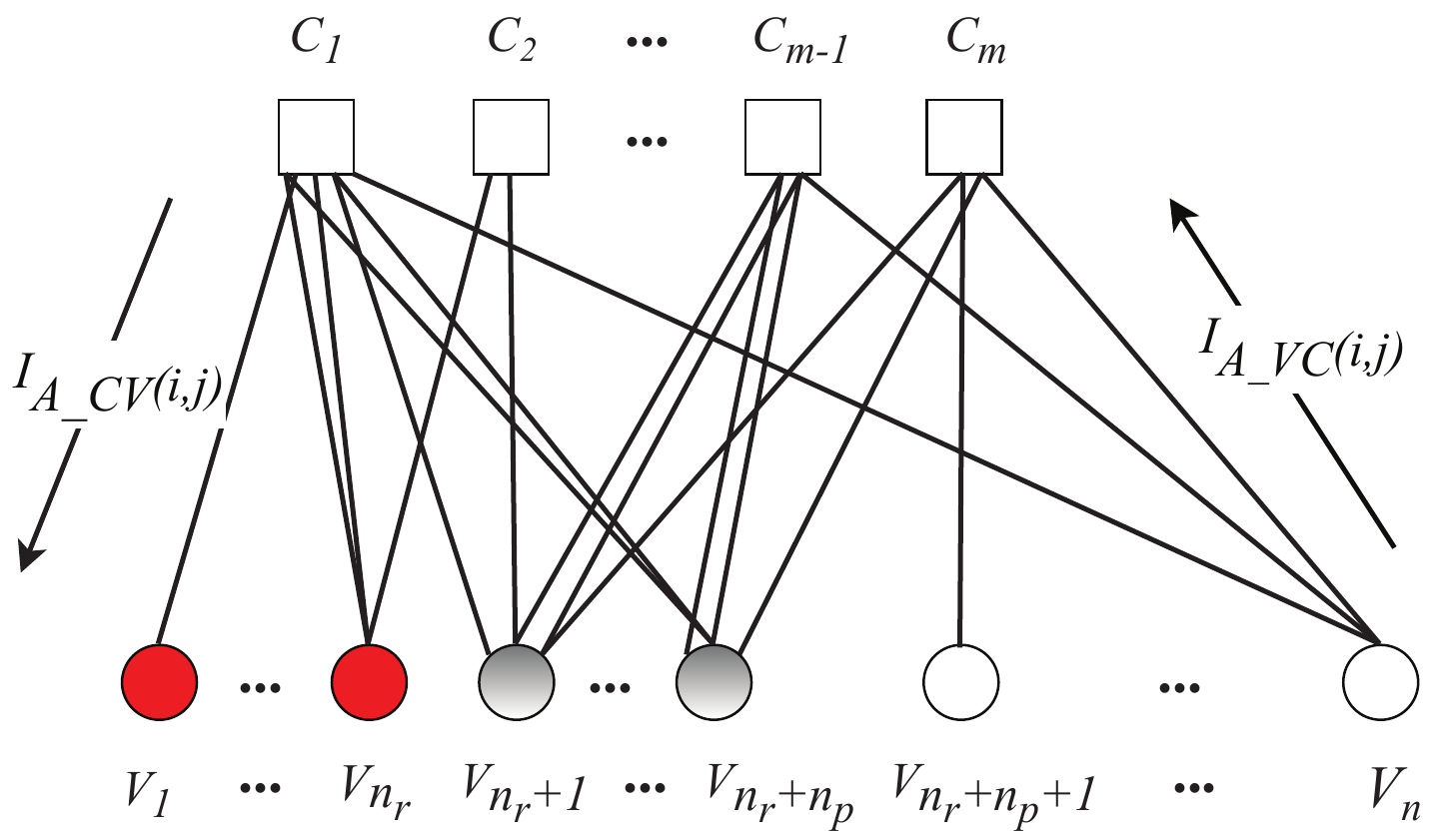}}
	\caption{The protograph of the PEXIT-JSCC algorithm.}
	\label{PEXIT_JSCC}
\end{figure}

Denoting  $E_s$ as the average transmitted energy per source symbol and $N_0$ as the noise power spectral density,
the AWGN noise variance $\sigma^2$ and  $E_s/N_0$ is related by 
% we can calculate the mean square error of the Gaussian noise added to the channel $\sigma^{2}$ based on 
\begin{equation}
\frac{E_s}{N_0}=10\log_{10} \frac{1}{2 \sigma^{2} R \; }~\text{dB}.
\label{eq:Es_N0}
\end{equation}
We further define the following.
\begin{itemize}
	\item{$I_{A\_{VC}}(i,j)$ denotes the a priori mutual information (AMI) from the $j$-th VN to the $i$-th CN} (see Fig.~\ref{PEXIT_JSCC}).
	\item{$I_{A\_{CV}}(i,j)$ denotes the AMI from the $i$-th CN to the $j$-th VN  (see Fig.~\ref{PEXIT_JSCC}).}
	\item{$I_{E\_{CV}}(i,j)$ denotes the extrinsic mutual information (EMI) from the $i$-th CN to the $j$-th VN.}
	\item{$I_{E\_{VC}}(i,j)$ denotes the EMI from the $j$-th VN to $i$-th CN.}
	\item{$I_{APP}(j)$ denotes the mutual information (MI) between the a posterior log-likelihood-ratio (APP-LLR) of the $j$-th VN and its corresponding symbol.}
	\item $(E_s/N_0)^{*}$ denotes the channel threshold.
		\item{The mutual information (MI) between the VN $V_s$ corresponding to the source symbol and its corresponding $LLR_s$ is defined by \cite{fre2010joint} 
		\begin{equation}
			\begin{split}
				&J_{\rm BSC}(\mu,p_1)\\
				&=p_1\times I(V_s;\omega^{(p_1)})+(1-p_1)\times I(V_s;\omega^{(1-p_1)})
			\end{split}
		\end{equation}
		where $\mu$ signifies the average LLR value of the VN $V_s$, $\omega^{(p_1)}\sim N(\mu-LLR_s,2\mu)$, $\omega^{(1-p_1)}\sim N(\mu+LLR_s,2\mu)$. $I(a;b)$ denotes the MI between $a$ and $b$.}
	\item{An indicator function $\psi(x)$ is defined as
		\begin{equation}
			\psi(x) = \left\{
			\begin{array}{cc}
				0, & \textrm{if $x=0$} \\
				1, & \textrm{otherwise.}
			\end{array}
			\right.
	\end{equation}}
\end{itemize}

In \textbf{Algorithm \ref{alg3}}, we present our generalized algorithm,
namely protograph EXIT for JSCC algorithm (PEXIT-JSCC algorithm),
 for analyzing the channel threshold of the proposed P-JSCC. 
Note that our generalized algorithm is similar to those used in analyzing DP-LDPC codes in the literature, e.g., those in \cite{chen2016performance}. %chen2018joint
However, our algorithm is generalized in the sense that the protograph does not need to satisfy
any specific constraint, i.e., the requirement given
by \eqref{eq:constraint} does not exist.  
The maximum number of iterations $t_{max}$, step size $\Delta$, and
tolerance value $\delta$ used in \textbf{Algorithm \ref{alg3}}
are listed in Table~\ref{para}.
Using the PEXIT-JSCC algorithm, the channel thresholds of  
AR3A-JSCC and AR4JA-JSCC under $p_1=0.04$
 are found and listed in Table \ref{opt_tabl_cth}. 

\begin{table}[!t]
	\caption{The parameter settings of the PEXIT-JSCC and SGP-EXIT algorithms}	
	\centering
	\begin{tabular}{|c|c|c|c|c|}
		\hline
		\multirow{2}{*}{PEXIT-JSCC} & $\Delta$ &  $t_{max}$ & $\delta$ \\
		\cline{2-4}
		 & $0.001$~dB & $200$ & $10^{-6}$ \\
		\hline
		\multirow{2}{*}{SSP-JSCC} & $\hat{p}_1$ & $l_{max}$ & $\theta$ \\
		\cline{2-4}
		 & $0.001$ & $200$ & $10^{-6}$ \\
		\hline		
	\end{tabular}
	\label{para}
\end{table}

\begin{table*}[!ht]
\caption{{\color{black} The channel thresholds and TMDR values of different protomatrices given $p_1=0.04$. The Shannon limit is $-7.00$ {\upshape d}B}}	
\centering
\begin{tabular}{|c||c|c||c|c|c||c|c|}
\hline
~ & AR3A-JSCC & AR4JA-JSCC & $\textbf{B}_{sp\_{opt1}}$ & $\textbf{B}_{sp\_{opt2}}$ & $\textbf{B}_{sp\_{opt3}}$ & $\textbf{B}_{sp\_{opt1}}^{47}$ & $\textbf{B}_{sp\_{opt2}}^{47}$\\
\hline
$(E_s/N_0)^{*}$ (dB) & $-5.918$ & $-5.767$ & $-6.102$ & $-5.810$ & $-5.782$ & {\color{black}$-6.163$} & {\color{black}$-5.909$} \\
\hline
TMDR value & None & $0.017$ & None & None & $0.003$ & None &  $0.007$  \\
\hline		
\end{tabular}
\label{opt_tabl_cth}
\end{table*}

%And, some functions are defined as follows:
%\begin{enumerate}
%%	\item{The initial source log-likelihood-ratio (LLR) $LLR_s$ is given 
%%		\begin{equation}
%%			LLR_s = \ln((1-p_1)/p_1)
%%	\end{equation}}
%	\item{The mutual information (MI) between the VN $V_s$ corresponding to the source symbol and its corresponding $LLR_s$ is defined by \cite{fre2010joint} 
%		\begin{equation}
%			\begin{split}
%				&J_{\rm BSC}(\mu,p_1)\\
%				&=p_1\times I(V_s;\omega^{(p_1)})+(1-p_1)\times I(V_s;\omega^{(1-p_1)})
%			\end{split}
%		\end{equation}
%		where $\mu$ signifies the average LLR value of the VN $V_s$, $\omega^{(p_1)}\sim N(\mu-LLR_s,2\mu)$, $\omega^{(1-p_1)}\sim N(\mu+LLR_s,2\mu)$. $I(a;b)$ denotes the MI between $a$ and $b$.}
%	\item{The indicator function is given by
%		\begin{equation}
%			\psi(x) = \left\{
%			\begin{array}{cc}
%				0, & \textrm{if $x=0$} \\
%				1, & \textrm{otherwise.}
%			\end{array}
%			\right.
%	\end{equation}}
%\end{enumerate}

\begin{algorithm*}[h]
\caption{The PEXIT-JSCC algorithm. The definitions $J(\cdot)$ and $J^{-1}(\cdot)$ are given in \cite{ten2004design,ten2001convergence}.}\label{PEXIT-JSCC}
\begin{algorithmic}
\STATE
{\color{black} \STATE \hspace{0.5cm}$\text{Set the maximum number of iterations}~t_{max},~\text{step size}~\Delta,~
\text{and tolerance value}~\delta$.
\STATE \hspace{0.5cm}$\text{Set a sufficiently small}~E_s/N_0$.
\STATE \hspace{0.5cm}$ \textbf{for~} \text{a given}~E_s/N_0~\textbf{do}$
\STATE \hspace{1cm}$ \text{Set}~ t=1, I_{E\_{VC}}(i,j)=I_{A\_{CV}}(i,j)=I_{E\_{CV}}(i,j)=I_{A\_{VC}}(i,j)=0~\text{and}~I_{APP}(j)=0, \forall i,j$
\STATE \hspace{1cm}$ \textbf{while}~\sum\limits_{j=1}^{n} (1-I_{APP}(j))>\delta~\textbf{and}~ t\leq t_{max}$
\STATE \hspace{1.5cm}$\textbf{for $i=1,2,..,m$, $j=1,2,..,n$ do}$
\STATE \hspace{2.0cm}$I_{E\_{VC}}(i,j)=\left\{
\begin{array}{cc}
\psi(e_{i,j})J_{\rm BSC}\Bigg(\sum\limits_{i^{'} \not\neq i}e_{i^{'},j}[J^{-1}(I_{A\_{CV}}(i^{'},j))]^2+(e_{i,j}-1)[J^{-1}(I_{A\_{CV}}(i,j))]^2,p_1\Bigg), \\
\textrm{$j=1,2,...,n_r$} \\
\psi(e_{i,j})J\Bigg(\sqrt{\sum\limits_{i^{'} \not\neq i}e_{i^{'},j}[J^{-1}(I_{A\_{CV}}(i^{'},j))]^2+(e_{i,j}-1)[J^{-1}(I_{A\_{CV}}(i,j))]^2+\sigma_{ch}^2(j)}\Bigg), \\ \textrm{$j=n_r+1,n_r+2,...,n$}
\end{array}
\right.$
\STATE \hspace{2.0cm}$\text{where}~\sigma_{ch}^2(j)=\left\{
\begin{array}{cc}
0 & j=n_r+1,n_r+2,...n_r+n_p, \\
4/\sigma^2 & j=n_r+n_p+1,n_r+n_p+2,...,n\\
\end{array}
\right.$	
\STATE \hspace{1.5cm}$\textbf{end for}$		
\STATE \hspace{1.5cm}$\text{Set}~I_{A\_{VC}}(i,j)=I_{E\_{VC}}(i,j),\forall i,j$
\STATE \hspace{1.5cm}$\textbf{for $i=1,2,..,m$, $j=1,2,..,n$ do}$
\STATE \hspace{2.0cm}$I_{E\_{CV}}(i,j)=\psi(e_{i,j})\Bigg(1-J\Bigg(\sqrt{\sum\limits_{j^{'} \not\neq j}
e_{i,j^{'}}[J^{-1}(1-I_{A\_{VC}}(i,j^{'}))]^2+(e_{i,j}-1)[J^{-1}(1-I_{A\_{VC}}(i,j))]^2}\Bigg)\Bigg)$
\STATE \hspace{1.5cm}$\textbf{end for}$
\STATE \hspace{1.5cm}$\text{Set}~I_{A\_{CV}}(i,j)=I_{E\_{CV}}(i,j),\forall i,j$	
\STATE \hspace{1.5cm}$\textbf{for $j=1,2,..,n$ do}$		
\STATE \hspace{2.0cm}$I_{APP}(j)=\left\{
\begin{array}{cc}
J_{\rm BSC}\Bigg(\sum\limits_{i}e_{i,j}[J^{-1}(I_{A\_{CV}}(i,j))]^{2},p_1\Bigg), & \textrm{$j=1,2,...,n_r$} \\
J\Bigg(\sqrt{\sum\limits_{i}e_{i,j}[J^{-1}(I_{A\_{CV}}(i,j))]^{2}+\sigma_{ch}^2(j)}\Bigg), & \textrm{$j=n_r+1,n_r+2,...,n$}
\end{array}
\right.$
\STATE \hspace{2.0cm}$\text{where}~\sigma_{ch}^2(j)=\left\{
\begin{array}{cc}
0 & j=n_r+1,n_r+2,...n_r+n_p, \\
4/\sigma^2 & j=n_r+n_p+1,n_r+n_p+2,...,n\\
\end{array}
\right.$	
\STATE \hspace{1.5cm}$\textbf{end for}$
\STATE \hspace{1.5cm}$\text{Set}~t=t+1$	 
\STATE \hspace{1cm}$\textbf{end while}$
\STATE \hspace{1cm}$\textbf{if}~\sum\limits_{j=1}^{n} (1-I_{APP}(j))<\delta~\textbf{then}$
\STATE \hspace{1.5cm}$(E_s/N_0)^*=E_s/N_0$
\STATE \hspace{1.5cm}$break$
\STATE \hspace{1cm}$\textbf{else}$
\STATE \hspace{1.5cm}$E_s/N_0=E_s/N_0+\Delta$
\STATE \hspace{1cm}$\textbf{end if}$
\STATE \hspace{0.5cm}$\textbf{end for}$	}
\end{algorithmic}
\label{alg3}
\end{algorithm*}

\begin{algorithm*}[!h]
\caption{Inner-code curve and outer-code curve.}\label{SSP_EXIT chart}
\begin{algorithmic}
\STATE 
\STATE {\textsc{Inner-code curve}}
{\color{black}
\STATE \hspace{0.5cm}$ \text{Given}~I_{A\_{CV_p}}(i,j) \in [0,1], i=1,2,...,m,j=1,2,...,n_r+n_p$
\STATE \hspace{0.5cm}$ \textbf{for}~i=1,2,...,m,~j=1,2,...,n_r~\textbf{do}$
\STATE \hspace{0.5cm}\begin{equation} \label{SGP-EXIT1}
I_{E\_{V_{p}C}}(i,j)=\psi(e_{i,j})J_{\rm BSC}\Bigg(\sum\limits_{i^{'} \not\neq i}e_{i^{'},j}[J^{-1}(I_{A\_{CV_p}}(i^{'},j))]^2+(e_{i,j}-1)[J^{-1}(I_{A\_{CV_p}}(i,j))]^2,p_1\Bigg)
\end{equation}	
\STATE \hspace{0.5cm}$ \textbf{end for}$	
\STATE \hspace{0.5cm}$ \textbf{for}~i=1,2,...,m,~j=n_r+1,n_r+2,...,n_r+n_p~\textbf{do}$
\STATE \hspace{0.5cm}\begin{equation} \label{SGP-EXIT2}
I_{E\_{V_{p}C}}(i,j)=\psi(e_{i,j})J\Bigg(\sqrt{\sum\limits_{i^{'} \not\neq i}e_{i^{'},j}[J^{-1}(I_{A\_{CV_p}}(i^{'},j))]^2+(e_{i,j}-1)[J^{-1}(I_{A\_{CV_p}}(i,j))]^2}\Bigg)
\end{equation}	
\STATE \hspace{0.5cm}$ \textbf{end for}$
\STATE {\textsc{Outer-code curve}}
\STATE \hspace{0.5cm}$ \text{Given}~I_{A\_{V_pC}}(i,j) \in [0,1]~\text{and}~I_{A\_{V_tC}}(i,j^{*})=1, i=1,2,...,m,j=1,2,...,n_r+n_p, j^{*}=n_r+n_p+1,...,n$
\STATE \hspace{0.5cm}$ \textbf{for}~i=1,2,...,m,~j=1,2,...,n_r+n_p~\textbf{do}$
\STATE \hspace{1.0cm}$I_{E\_{CV_{p}}}(i,j)=\psi(e_{i,j})\Bigg(1-J\Bigg(\sqrt{\sum\limits_{j^{'} \not\neq j}
e_{i,j^{'}}[J^{-1}(1-I_{A\_{V_pC}}(i,j^{'}))]^2
+\sum\limits_{j^{*}=n_r+n_p+1}^{n}e_{i,j^{*}}[J^{-1}(1-I_{A\_{V_tC}}(i,j^{*}))]^2}$
\STATE \hspace{12.0cm}$\overline{+(e_{i,j}-1)[J^{-1}(1-I_{A\_{V_pC}}(i,j))]^2}\Bigg)\Bigg)$	
\STATE \hspace{1.0cm}\begin{equation} \label{SGP-EXIT3}
=\psi(e_{i,j})\Bigg(1-J\Bigg(\sqrt{\sum\limits_{j^{'} \not\neq j}
e_{i,j^{'}}[J^{-1}(1-I_{A\_{V_pC}}(i,j^{'}))]^2+(e_{i,j}-1)[J^{-1}(1-I_{A\_{V_pC}}(i,j))]^2}\Bigg)\Bigg)\end{equation}
\STATE \hspace{0.5cm}$ \textbf{end for}$}
%	\STATE The definitions $J(\cdot)$ and $J^{-1}(\cdot)$ are given in \cite{ten2004design,ten2001convergence}.	  
\end{algorithmic}
\label{alg1}
\end{algorithm*}

\begin{figure*}[!t]
	\centering
	\subfloat[]{\includegraphics[width=4.0in]{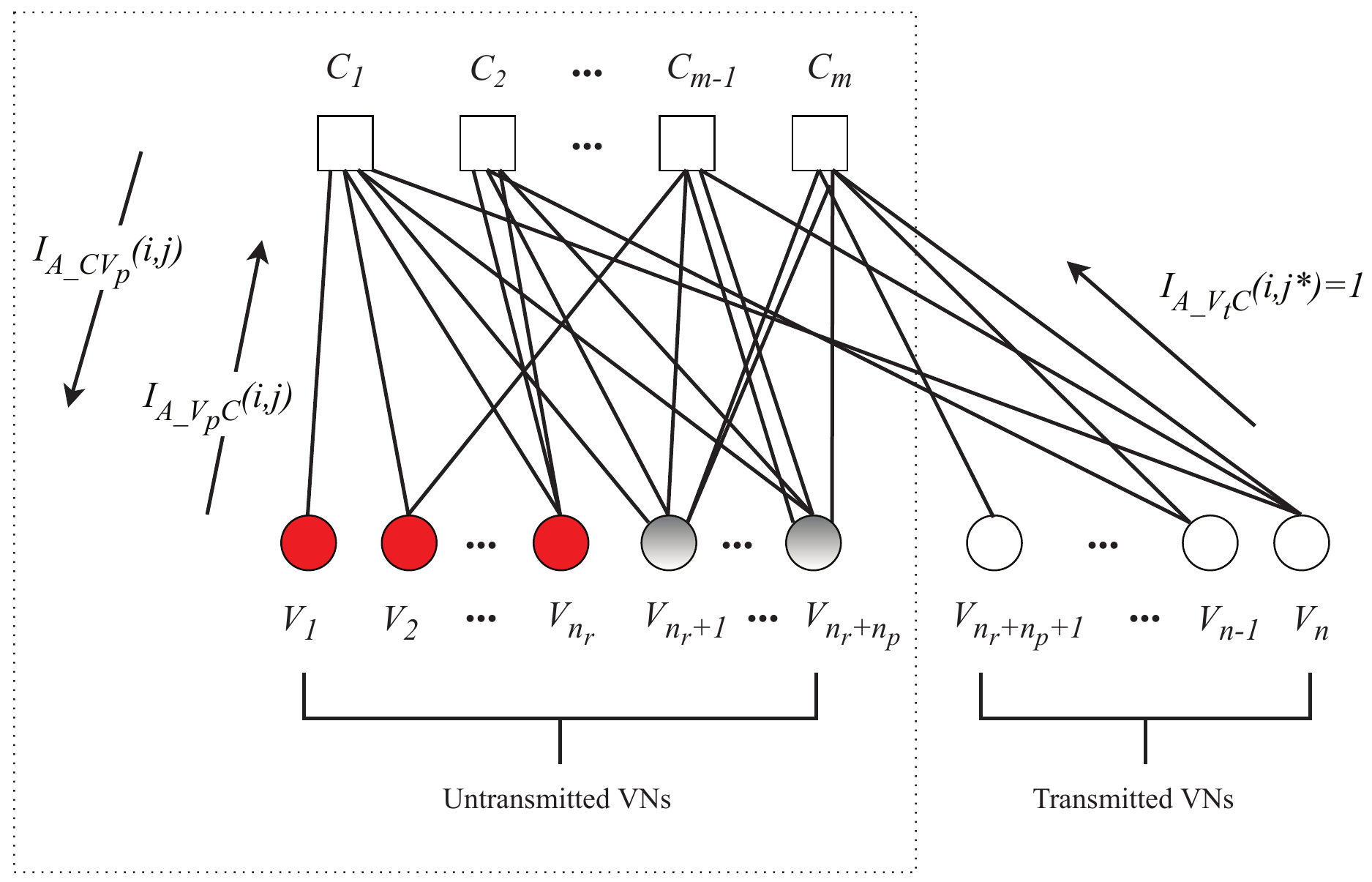}
		\label{SGP-EXIT_a}}
	\hfil
	\subfloat[]{\includegraphics[width=4.5in]{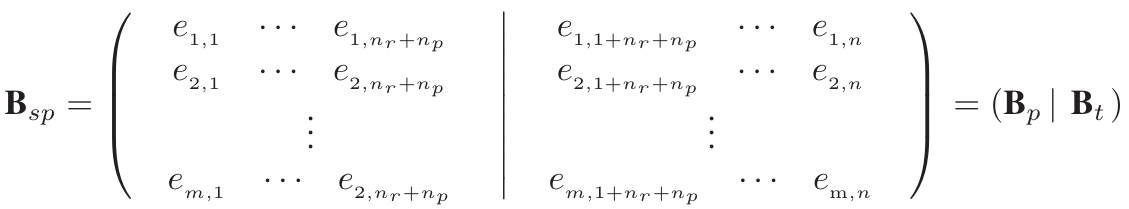}
		\label{SGP-EXIT_b}}
	\caption{(a) Passing of AMI in the protograph representing a P-JSCC. (b) The  protomatrix of a P-JSCC is split into two sub-protomatrices. $\textbf{B}_{p}$ contains the untransmitted VNs and $\textbf{B}_{t}$ contains the transmitted VNs.}
	\label{SGP-EXIT}
\end{figure*}

\subsection{SGP-EXIT chart}
The algorithms in \cite{chen2015matching,chen2020analysis} are not suitable for calculating the source thresholds of the P-JSCC system in Fig.~\ref{JSCC_SPLDPC}. In \cite{chen2015matching}, the SPEXIT algorithm is applied to calculate the source threshold of a double protograph with no connections between VNs in the source P-LDPC code and CNs in the channel P-LDPC code. In \cite{chen2020analysis}, the ESP-EXIT algorithm does not consider the case with punctured variable nodes. Also, there is a constraint on the structure of $\textbf{B}_{svcc}$, i.e., each non-zero column only allows a weight of 2. Here, we propose a more generic algorithm for calculating the source threshold of a JSCC system, called the source generic protograph EXIT (SGP-EXIT) algorithm. 

First, we refer to Fig.~\ref{SGP-EXIT}(a) and define the following.
\begin{itemize}
\item $\mathcal{V}_s=\{V_1, V_2, ..., V_{n_r}\}$ denotes the set of VNs corresponding to source symbols.
\item $\mathcal{V}_p=\{V_1, V_2, ..., V_{n_r+n_p}\}$ denotes the set of untransmitted VNs of the generic protograph.
\item $\mathcal{V}_t=\{V_{n_r+n_p+1}, V_{n_r+n_p+2}, ..., V_{n}\}$ denotes the set of transmitted VNs.
\item ${\mathcal{C}}= \{C_1,C_2,...,C_m\}$ denotes the set of CNs.
\item $I_{A\_{V_pC}}(i,j)$  denotes the AMI from the $j$-th untransmitted VN $\in \mathcal{V}_p$ to the $i$-th CN.
\item $I_{A\_{V_tC}}(i,j^{*})$ denotes the AMI from the $j^{*}$-th transmitted VN $\in \mathcal{V}_t$ to the $i$-th CN.
\item{$I_{A\_{CV_p}}(i,j)$ denotes the AMI from the $i$-th CN $\in{\mathcal{C}}$ to the $j$-th untransmitted VN $\in \mathcal{V}_p$.}
	\item{$I_{E\_{CV_{p}}}(i,j)$ denotes the EMI from the $i$-th CN $\in{\mathcal{C}}$ to the $j$-th untransmitted VN $\in \mathcal{V}_p$.}
	\item{$I_{E\_{V_{p}C}}(i,j)$ denotes the EMI from the $j$-th untransmitted VN $\in \mathcal{V}_p$ to the $i$-th CN.}
	\item{$I_{APP}\_p(j)$ denotes the MI between the APP-LLR of the $j$-th untransmitted VN $\in \mathcal{V}_p$ and its corresponding symbol.}
\end{itemize}
Next, we can derive the SGP-EXIT  curves 
(i.e., inner-code curve and outer-code curve) 
of a given P-JSCC using \textbf{Algorithm~\ref{alg1}}.
Note that our objective is to investigate the effect of the source probability $p_1$ on 
the source symbol error performance 
at the high SNR region, i.e., when the noise power is very small.
Thus we assume that the average AMI from the transmitted VNs to CNs is equal to $1$, i.e., $I_{A\_{V_tC}}(i,j^{*})=1$ and $J^{-1}(1-I_{A\_{V_tC}}(i,j^{*}))=0$ for $i=1,2,...,m,j^{*}=n_r+n_p+1,n_r+n_p+2,...,n$, and apply
it in \eqref{SGP-EXIT3} in \textbf{Algorithm~\ref{alg1}}. 
As can be observed in  \textbf{Algorithm~\ref{alg1}} and Fig.~\ref{SGP-EXIT}(a),
 we only need to consider the untransmitted VNs $\mathcal{V}_p$ and their connected CNs
 in deriving the SGP-EXIT curves. 
 In other words, we only need to consider the sub-protomatrix
\begin{equation}
\textbf{B}_p=\left(\begin{array}{*{20}{c}}
		{\begin{array}{*{20}{c}}
				{{e_{1,1}}} & {...} & {{e_{1,{n_r}}}} & {...} & {{e_{1,{n_r+n_p}}}}   \\
		\end{array}}  \\
		{\begin{array}{*{20}{c}}
				{{e_{2,1}}} & {...} & {{e_{2,{n_r}}}} & {...} & {{e_{2,{n_r+n_p}}}}  \\
		\end{array}}  \\
		\vdots   \\
		{\begin{array}{*{20}{c}}
				{{e_{m,1}}} & {...} & {{e_{m,{n_r}}}} & {...} & {{e_{m,{n_r+n_p}}}}   \\
		\end{array}}  \\
\end{array}\right)
\end{equation}
which is also shown in Fig.~\ref{SGP-EXIT}\subref{SGP-EXIT_b}. 

\begin{figure}[!t]
	\centering
	\includegraphics[width=3.2in]{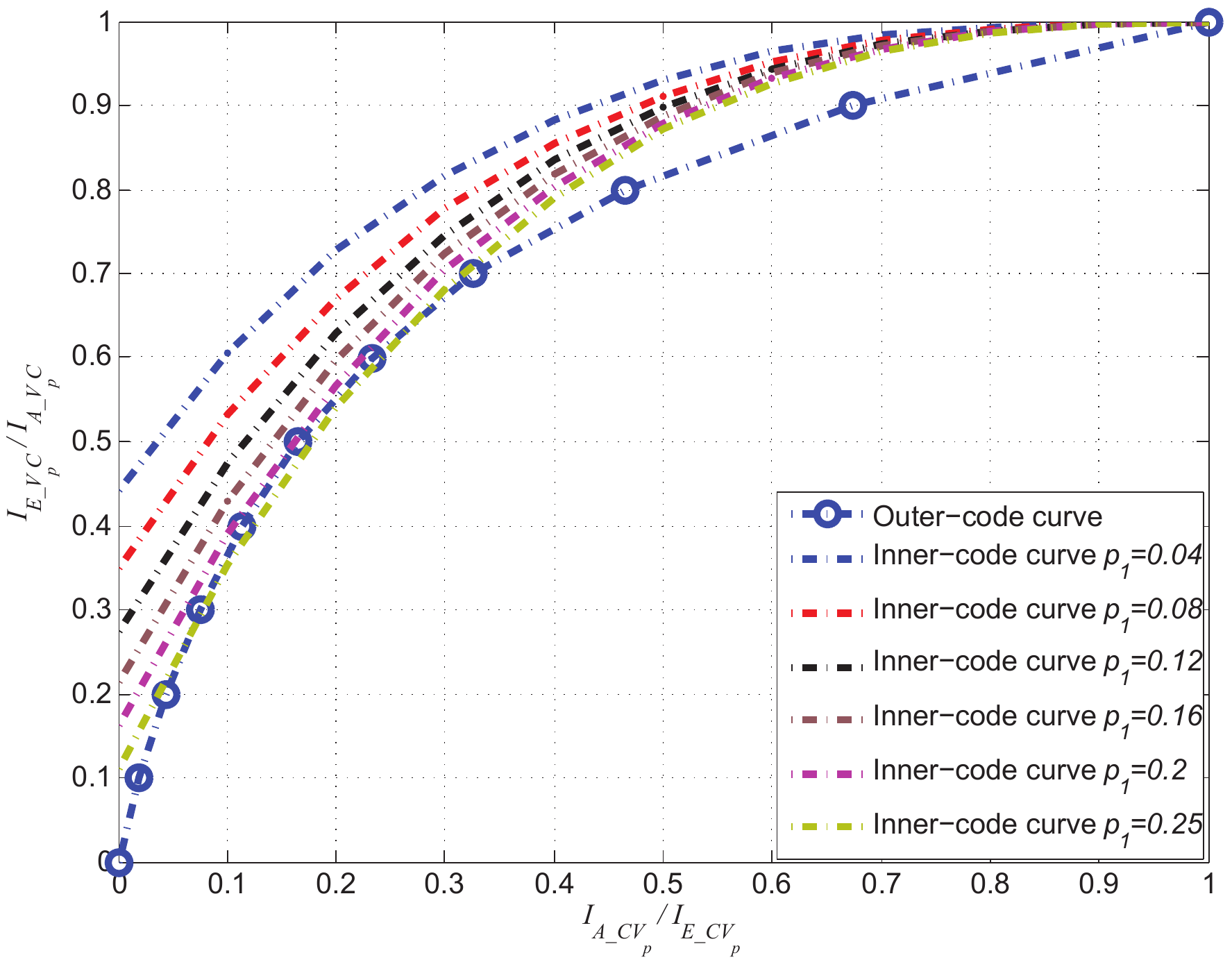}
	\caption{The SGP-EXIT chart of the sub-protomatrix $\textbf{B}_{p}$ in  \eqref{sub-opt1} under different $p_1$ values. Estimated source threshold is $0.25$.}
	\label{ssp_chart}
\end{figure}

\begin{figure}[!t]
	\centering
	\includegraphics[width=3.2in]{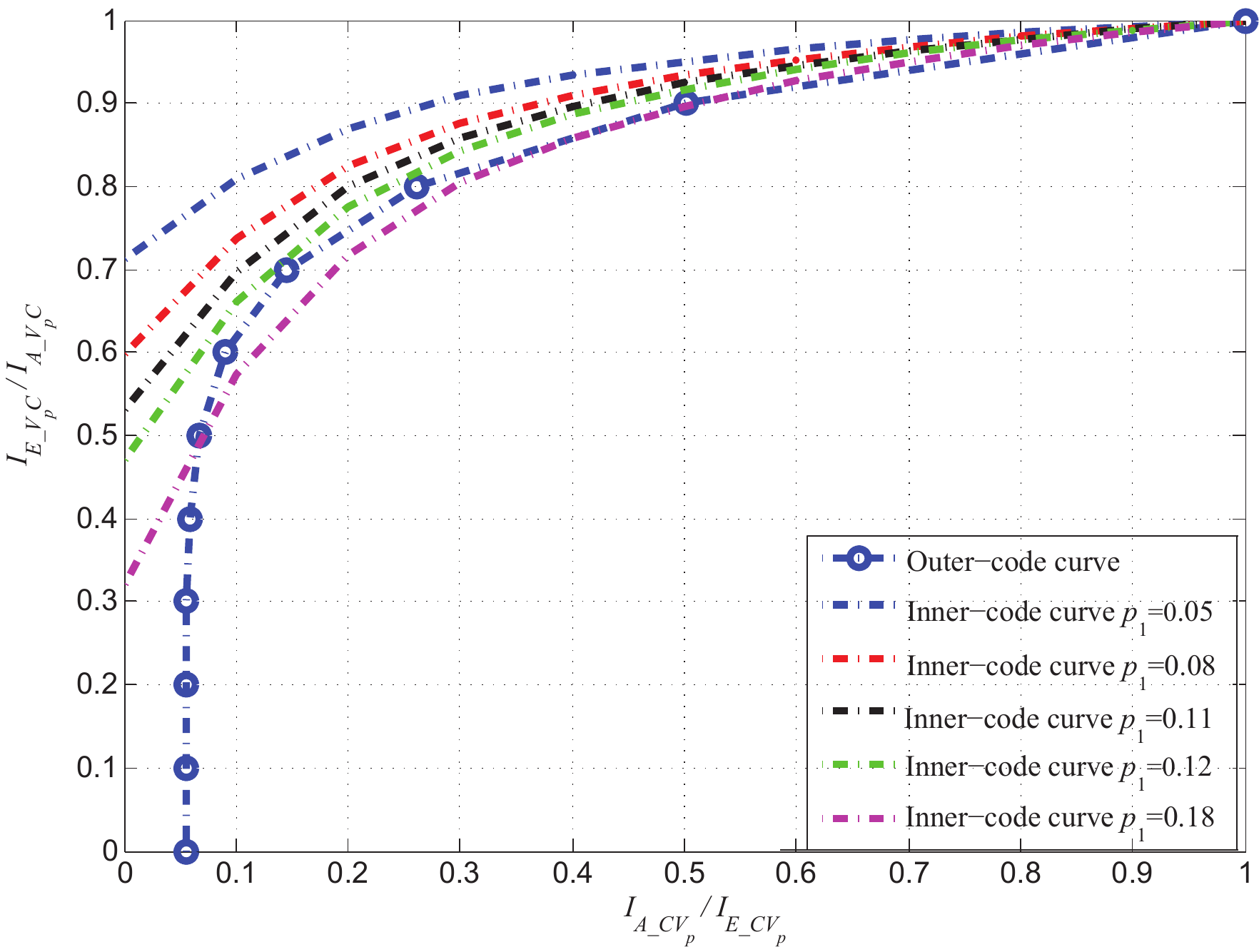}
	\caption{The SGP-EXIT chart of the sub-protomatrix $\textbf{B}_{p}$ in (\ref{BJ}) under different $p_1$ values. Estimated source threshold is $0.12$.}
	\label{exit_chart}
\end{figure}

Considering the protomatrix 
\begin{equation} \label{opt1_0}
	\begin{array}{l}
		{\textbf{B}}_{sp\_{opt1}}
		= \left( {\begin{array}{*{20}{c}}
				1 & 1 & 1 & 1 & 1 \\
				0 & 0 & 2 & 0 & 1 \\
				3 & 2 & 2 & 0 & 0 \\
		\end{array}} \right) \\ 
	\end{array}
\end{equation}
and assuming $n_r =2$ and $n_p=1$, 
we form the sub-protomatrix
\begin{equation} \label{sub-opt1}
	\begin{array}{l}
		{\textbf{B}}_{p}
		= \left( {\begin{array}{*{20}{c}}
				1 & 1 & 1  \\
				0 & 0 & 2  \\
				3 & 2 & 2  \\
		\end{array}} \right) \\ 
	\end{array}
\end{equation}
and plot the corresponding SGP-EXIT curves in
Fig.~\ref{ssp_chart}. 
We can see that the gap between the inner-coder curve and the outer-code curve becomes smaller as $p_1$ increases. When $p_1$ increases beyond a certain value, these two curves will cross each other. The decoding can be performed successfully only when the inner-coder curve is above the outer-coder curve. The larger the gap between the outer-code curve and the inner-code curve, the faster the decoder converges. The maximum $p_1$ value that makes these two curves closest without crossing is the source threshold $p_{1\_th}$. We can see from Fig.~\ref{exit_chart} that when $p_1=0.25$, the two curves are closest. Then the source threshold of 
the sub-protomatrix $\textbf{B}_{p}$ in  \eqref{sub-opt1} is estimated to be $p_{1\_th}=0.25$.
To achieve a more precise source threshold, we propose the SGP-EXIT algorithm shown in \textbf{Algorithm~\ref{alg2}}. 
The maximum number of iterations $l_{max}$, step size $\hat{p}_1$
and tolerance value $\theta$ used in the algorithm are listed  
 in Table~\ref{para}. 
Using these parameters, we obtain also 
$p_{1\_th}=0.25$ for the sub-protomatrix $\textbf{B}_{p}$ in  \eqref{sub-opt1}.
Similarly, we apply \textbf{Algorithm~\ref{alg2}} to obtain the source thresholds of the AR3A-JSCC code and AR4JA-JSCC code in 
	Fig.~\ref{fig:matrices}. The source thresholds of the above codes are listed 	
 in Table~\ref{opt_tabl_sth}.

\begin{table*}[!t]
	\caption{The source thresholds of different protomatrices.}	
	\centering
	\begin{tabular}{|c||c|c||c|c|c||c|c|}
		\hline
		~ & AR3A-JSCC & AR4JA-JSCC & $\textbf{B}_{sp\_{opt1}}$ & $\textbf{B}_{sp\_{opt2}}$ & $\textbf{B}_{sp\_{opt3}}$  & $\textbf{B}_{sp\_{opt1}}^{47}$ & $\textbf{B}_{sp\_{opt2}}^{47}$\\
		\hline
		$p_{1\_{th}}$ & $0.228$ & $0.212$ & $0.25$ & $0.275$ & $0.242$ & {\color{black}$0.290$} & {\color{black}$0.276$}  \\
		\hline		
	\end{tabular}
	\label{opt_tabl_sth}
\end{table*}

\begin{algorithm*}[!t]
\caption{SGP-EXIT algorithm.}\label{SSP_EXIT value}
\begin{algorithmic}
\STATE
{\color{black}
\STATE \hspace{0.5cm}$\text{Set the maximum number of iterations}~l_{max},~\text{step size}~\hat{p}_1~
\text{and tolerance value}~\theta$. 
{\color{black}\STATE \hspace{0.5cm}$ \text{Set a sufficiently large}~ p_1<0.5$.
\STATE \hspace{0.5cm}$ \textbf{for~} \text{a given}~ p_1<0.5~\textbf{do}$}
\STATE \hspace{1cm}$ \text{Set}~ l=1, I_{E\_{V_{p}C}}(i,j)=I_{A\_{CV_p}}(i,j)=I_{E\_{CV_{p}}}(i,j)=I_{A\_{V_pC}}(i,j)=0, I_{APP}\_p(j)=0, i=1,2,..,m, j=$
\STATE \hspace{1cm}$1,2,...,n_r+n_p$.
\STATE \hspace{1cm}$ \textbf{while}~\sum\limits_{j=1}^{n_r+n_p} (1-I_{APP}\_p(j))>\theta~\textbf{and}~ l\leq l_{max}$
\STATE \hspace{1.5cm}$\text{Compute}~I_{E\_{V_{p}C}}(i,j)~\text{using (\ref{SGP-EXIT1}) and (\ref{SGP-EXIT2})} \ \forall i, j$.			
\STATE \hspace{1.5cm}$\text{Set}~I_{A\_{V_{p}C}}(i,j)=I_{E\_{V_{p}C}}(i,j) \ \forall i,j$.
\STATE \hspace{1.5cm}$\text{Compute}~I_{E\_{CV_{p}}}(i,j)~\text{using (\ref{SGP-EXIT3})} \ \forall i,j$.
\STATE \hspace{1.5cm}$\text{Set}~I_{A\_{CV_{p}}}(i,j)=I_{E\_{CV_{p}}}(i,j) \ \forall i,j$.	
\STATE \hspace{1.5cm}$ \text{Calculate}~I_{APP}\_p(j)=\left\{
\begin{array}{cc}
J_{\rm BSC}\Bigg(\sum\limits_{i}e_{i,j}[J^{-1}(I_{A\_{CV_p}}(i,j))]^{2},p_1\Bigg), & \textrm{$j=1,2,...,n_r$} \\
J\Bigg(\sqrt{\sum\limits_{i}e_{i,j}[J^{-1}(I_{A\_{CV_p}}(i,j))]^{2}}\Bigg), & \textrm{$j=n_r+1,n_r+2,...,n_r+n_p$}
\end{array}
\right. .$	 
\STATE \hspace{1.5cm}$\text{Set}~l=l+1$.
\STATE \hspace{1cm}$\textbf{end while}$
\STATE \hspace{1cm}$\textbf{if}~\sum\limits_{j=1}^{n_r+n_p} (1-I_{APP}\_p(j))<\theta~\textbf{then}$
\STATE \hspace{1.5cm}$p_{1\_{th}}=p_1$
\STATE \hspace{1.5cm}$break$
\STATE \hspace{1cm}$\textbf{else}$
\STATE \hspace{1.5cm}$p_1=p_1-\hat{p}_1$
\STATE \hspace{1cm}$\textbf{end if}$
\STATE \hspace{0.5cm}$\textbf{end for}$}
\end{algorithmic}
\label{alg2}
\end{algorithm*}

%
%\begin{equation}\label{BJ}
%	\begin{array}{l}
%		{{\textbf{B}}_p} = \left( {\begin{array}{*{20}{c}}
%				{{\textbf{B}_s}}  \\
%				{{\textbf{B}_{svcc}}}  \\
%		\end{array}} \right)  
%		= \left( {\begin{array}{*{20}{c}}
%				3 & 2 & 1 & 1 & 0 & 1 & 0 & 0  \\
%				2 & 3 & 1 & 0 & 1 & 0 & 1 & 0  \\
%				3 & 3 & 0 & 0 & 0 & 0 & 0 & 1 \\
%				3 & 0 & 1 & 2 & 2 & 1 & 1 & 1  \\
%				0 & 0 & 0 & 0 & 0 & 0 & 0 & 0  \\
%				0 & 0 & 1 & 0 & 0 & 0 & 0 & 0  \\
%				0 & 0 & 0 & 0 & 0 & 0 & 0 & 0  \\
%				0 & 0 & 1 & 0 & 0 & 0 & 0 & 0  \\
%		\end{array}} \right) \\ 
%	\end{array}
%\end{equation}
%where $n_p=0, n_r=8, m=8$. 

We further consider a DP-LDPC code given by \cite[Eq.(26)]{chen2020analysis} 
where $m=8, n=16, n_r=8, n_p=0$. Since there is no punctured VN, the 
sub-protomatrix needs to be considered when deriving the source threshold is
simply the sub-protomatrix 
$\left( {\begin{array}{*{20}{c}}
		{{\textbf{B}_s}}  \\
		{{\textbf{B}_{svcc}}}  \\
\end{array}} \right)$.  
The sub-protomatrix being studied is given by  \cite[Eq.(26)]{chen2020analysis} 
\begin{equation}\label{BJ}
	\begin{array}{l}
		{{\textbf{B}}_p} = \left( {\begin{array}{*{20}{c}}
				{{\textbf{B}_s}}  \\
				{{\textbf{B}_{svcc}}}  \\
		\end{array}} \right)  
		= \left( {\begin{array}{*{20}{c}}
				3 & 2 & 1 & 1 & 0 & 1 & 0 & 0  \\
				2 & 3 & 1 & 0 & 1 & 0 & 1 & 0  \\
				3 & 3 & 0 & 0 & 0 & 0 & 0 & 1 \\
				3 & 0 & 1 & 2 & 2 & 1 & 1 & 1  \\
				0 & 0 & 0 & 0 & 0 & 0 & 0 & 0  \\
				0 & 0 & 1 & 0 & 0 & 0 & 0 & 0  \\
				0 & 0 & 0 & 0 & 0 & 0 & 0 & 0  \\
				0 & 0 & 1 & 0 & 0 & 0 & 0 & 0  \\
		\end{array}} \right) \\ 
	\end{array}.
\end{equation}
Fig.~\ref{exit_chart} plots the SGP-EXIT chart using \textbf{Algorithm~\ref{alg1}} under different $p_1$ values. The estimated source threshold is $0.12$. 
Using  \textbf{Algorithm~\ref{alg2}}, we obtain a more precise threshold, i.e.,  $p_{1\_{th}}=0.1156$. This result is the same as that calculated by the ESP-EXIT algorithm in \cite{chen2020analysis}.

\subsection{Optimization method}
We aim to design P-JSCCs with good error performance in both 
 waterfall and high-SNR regions. 
Our objective is therefore to construct generic protographs with a high $p_{1\_{th}}$ based on the SGP-EXIT algorithm and a low channel decoding threshold $(E_s/N_0)^{*}$ based on the PEXIT-JSCC algorithm. Moreover, we use the AWD tool to analyze the linear minimum distance properties of the constructed generic protographs.

\begin{figure*}[!t]
	\centering
	\includegraphics[width=6.0in]{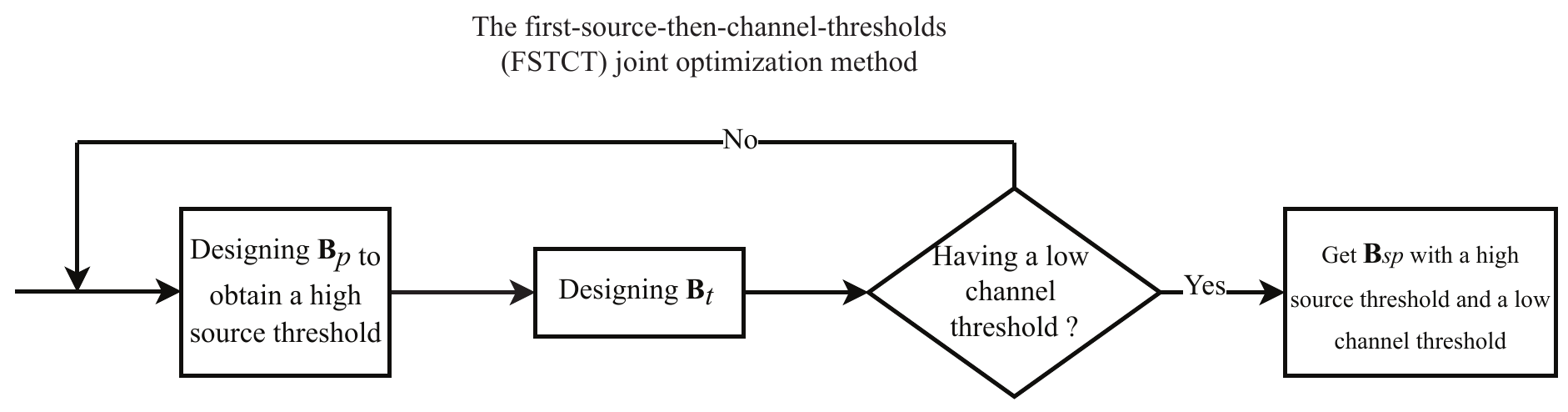}
	\caption{The flow of the optimization algorithm.}
	\label{fig_method}
\end{figure*}

We propose a first-source-then-channel-thresholds (FSTCT) joint optimization method to design generic protographs. The method is implemented in two steps. 
The first one is to optimize the connections between the untransmitted VNs and the CNs, i.e., $\textbf{B}_{p}$ shown in Fig.~\ref{SGP-EXIT}\subref{SGP-EXIT_b}, to obtain a high source threshold (based on the SGP-EXIT algorithm). 
The second one is to design the remaining part of $\textbf{B}_{sp}$, i.e., $\textbf{B}_{t}$ shown in Fig.~\ref{SGP-EXIT}\subref{SGP-EXIT_b}, for a given $p_1$ to achieve a low channel threshold (based on the PEXIT-JSCC algorithm). 
The search scope is significantly reduced because $\textbf{B}_{sp}$ is divided into two parts and are to be optimized one after another. 
The flow of this method is shown in Fig.~\ref{fig_method}. By using this optimization method, we can guarantee that the obtained generic protographs have good source and channel thresholds.

As an illustration, we assume a symbol code rate of $R=1$ and $n_p=1$ punctured VN. 
In order to reduce the search space
and based on some existing code design rules, we further assume the following conditions.
\begin{enumerate}[(i)]
\item \label{con1} The maximum value of each entry in $\textbf{B}_{sp}$ is $e_{max}=3$.
%\item \label{con2} 
\item The minimum size of the generic protomatrix is $3\times 5$. For $k\in Z^{+}$, the number of VNs corresponding to source symbols is $n_r=k+1$. According to 
 (\ref{rate}), the total number of VNs is $n=2k+3$ and the total number of CNs is $m=n-n_r=k+2$.
	\item \label{con3} The degree of each CN in $\textbf{B}_{sp}$ is at least 3. 
	\item \label{con4} The maximum number of degree-2 VNs in $\textbf{B}_{sp}$ is $n-m-n_p=k$. 
	\item \label{con5} The maximum number of degree-1 VNs in $\textbf{B}_{sp}$ is $1$ and $n_p$ VNs with the highest degrees in $\textbf{B}_{sp}$ are punctured.
	\item \label{con6} The maximum degree of each VN in $\textbf{B}_{sp}$ is $D_{max}$. (In this paper, we set $D_{max}=8$ when $k=1$; and $D_{max}=11$ when $k=2$.)
	\item \label{con7} $p_{1\_{th}}> \bar{p}_1$ where $\bar{p}_1$ is a preset source threshold benchmark.
	\item \label{con8} $(E_s/N_0)^{*}<(E_s/N_0){'}$ where $(E_s/N_0){'}$ is a preset channel threshold benchmark.
\end{enumerate}

When the search scope is small, a brute-force approach can be used to 
search for good generic protographs. 
For example 
when $k=1$, the size of $\textbf{B}_{p}$ is $3\times 3$ and the size of $\textbf{B}_{t}$ is $3\times 2$. In this case, a brute-force search is an option. 
When the protomatrix is large, the brute-force approach will be very time consuming and even not feasible. Then a differential evolution (DE) strategy can be used. The following is a more in-depth explanation of these two search methods.

\subsubsection{Brute-force search}
To begin with, we do an exhaustive search for all entries in $\textbf{B}_p$ that fulfill  Conditions (\ref{con1}) to (\ref{con7}) in accordance with the SGP-EXIT algorithm. Among the ${\textbf{B}_{p}}$'s having high source thresholds, we exhaustively search through all entries in $\textbf{B}_{t}$ to make sure $\textbf{B}_{sp}$ satisfies the conditions (\ref{con1}) to (\ref{con6}) and (\ref{con8}) according to the PEXIT-JSCC algorithm. 

\subsubsection{DE approach}
We define
\begin{itemize}
	\item $G$ as the number of generations;
	\item $S$ as the number of candidate matrices;
	\item $p_c$ as the crossover probability.
\end{itemize}

\begin{enumerate}[Step a)] \label{conditions}
	\item \label{step1} Initialization: For the $0$-th generation, randomly generate candidate matrices $\textbf{B}_{bp\_1}^{0},\textbf{B}_{bp\_2}^{0},...,\textbf{B}_{bp\_S}^{0}$, 
the size of which equals that of $\textbf{B}_{p}$ or $\textbf{B}_{t}$, depending on what matrix is to be constructed.
Moreover, these matrices should satisfy Conditions (\ref{con1}) to (\ref{con6}). Set $g=0$.
	\item \label{step2} Mutation: Generate $S$ mutation matrices from Generation $g$ using 
	\begin{equation}
		\begin{split}
			\textbf{B}_{M\_s}^{g}&=\Phi(\textbf{B}_{bp\_{r1}}^{g}+0.5*(\textbf{B}_{bp\_{r2}}^{g}-\textbf{B}_{bp\_{r3}}^{g})), \\
			&\makebox[3cm]{} s=1,2,...,S
		\end{split}
	\end{equation}
	where $r1$, $r2$ and $r3$ are distinct random positive integers in the range $[1,S]$, and $\Phi(\cdot)$ is a function that converts each entry in a matrix
	to an integer closest to its absolute value.
	\item \label{step3} Crossover: 
	Create the matrix $\textbf{B}_{cr\_s}^{g}$ ($s=1,2,...,S$), in which
	the $(i,j)$-th element is set as the $(i,j)$-th element in $\textbf{B}_{M\_s}^{g}$ with  probability $p_c$, or as the $(i,j)$-th element in ${\textbf{B}}_{bp\_s}^{g}$ with  probability $1-p_c$.
	\item \label{step4} Selection: Generate the $(g+1)$-th generation candidate matrices $\textbf{B}_{bp\_s}^{g+1}$  ($s=1,2,...,S$). Two cases are to be discussed.
	\begin{enumerate}
	  \item Case One: When the goal is to find a $\textbf{B}_{p}$ with a high source threshold, $\textbf{B}_{bp\_s}^{g+1}$  ($s=1,2,...,S$) is generated as
		\begin{equation}
			\textbf{B}_{bp\_s}^{g+1} = \left\{
			\begin{array}{cc}
%				\textbf{B}_{cr\_s}^{g}, & \textrm{if $\Theta(\textbf{B}_{cr\_s}^{g})\Psi(\textbf{B}_{cr\_s}^{g})$}\\
%				~ & \textrm{$> \Theta(\textbf{B}_{bp\_s}^{g})$}\\
				\textbf{B}_{cr\_s}^{g}, & \textrm{if $\Theta(\textbf{B}_{cr\_s}^{g})\Psi(\textbf{B}_{cr\_s}^{g})$} > \Theta(\textbf{B}_{bp\_s}^{g}) \\
				\textbf{B}_{bp\_s}^{g}, & \makebox[1cm]{}  \textrm{otherwise}
			\end{array}
			\right.
		\end{equation}
		where $\Theta(\textbf{B}_p)$ returns the source threshold $p_{1\_th}$ of $\textbf{B}_p$ according to the SGP-EXIT algorithm; and
		\begin{equation}
			\Psi(\textbf{B}_{cr\_s}^{g})=\left\{\begin{array}{cc}
				1, & \textrm{if Conditions (\ref{con1}) to (\ref{con6}) are } \\
				~  & \textrm{satisfied by $\textbf{B}_{cr\_s}^{g}$} \\
				0, & \textrm{otherwise.}
			\end{array}
			\right. 
		\end{equation}
	  \item Case Two: When the goal is to find a $\textbf{B}_{t}$ such that together with a pre-determined $\textbf{B}_p$ the generic protomatrix $\textbf{B}_{sp} = (\textbf{B}_p \ \textbf{B}_{t})$ can achieve a
	  low channel threshold $(E_s/N_0)^{*}$, $\textbf{B}_{bp\_s}^{g+1}$  ($s=1,2,...,S$) is generated as
	  \begin{equation}
	  	\textbf{B}_{bp\_s}^{g+1} = \left\{
	  	\begin{array}{cc}
	  		\textbf{B}_{cr\_s}^{g}, & \textrm{if~$\Upsilon\Big(\left(\textbf{B}_p~ \textbf{B}_{cr\_s}^{g}\right)\Big)\Psi(\textbf{B}_{cr\_s}^{g})$} \\	  	
	  		~ & \makebox[1.5cm]{}  \textrm{$< \Upsilon\Big((\textbf{B}_p~ \textbf{B}_{bp\_s}^{g})\Big)$}\\
	  		\textbf{B}_{bp\_s}^{g}, & \textrm{otherwise}
	  	\end{array}
	  	\right.
	  \end{equation}
	  where $\Upsilon(\textbf{B}_{sp})$ returns the channel threshold value 
	  of $\textbf{B}_{sp}$ according to the PEXIT-JSCC algorithm.
	  %, and $\textbf{B}_p$ is the  sub-protomatrix in $\textbf{B}_{sp}$ that has been pre-determined.
	\end{enumerate}
	\item Termination: Set $g=g+1$. Stop if $g=G$; otherwise go to Step \ref{step2}).
\end{enumerate}

\section{Results and discussions}\label{sect:results}
In this section, we present some optimized generic protomatrices, and their theoretical thresholds and error rate simulation results. 
As discussed in the previous section, the channel thresholds and source thresholds of 
the AR3A-JSCC and AR4JA-JSCC codes (see Fig.~\ref{fig:matrices})
have been derived  by the PEXIT-JSCC algorithm under $p_1=0.04$ and the SGP-EXIT algorithm, respectively. 
Referring to Tables \ref{opt_tabl_cth} and \ref{opt_tabl_sth},  
the channel thresholds of AR3A-JSCC and AR4JA-JSCC codes are, respectively, $-5.918$ dB and $-5.767$ dB;
and the source thresholds are, respectively, 
 $0.228$ and $0.212$.
Based on the results of the AR3A-JSCC and AR4JA-JSCC codes, 
we construct two sets of benchmarks, i.e., 
\begin{itemize}
\item $B_1: (\bar{p}_1=0.228, (E_s/N_0)'=-5.918 \ \text{dB})$; and
\item $B_2: (\bar{p}_1=0.212, (E_s/N_0)'=-5.767 \ \text{dB})$. 
\end{itemize}

%
%To obtain a generic protograph with both better source and channel thresholds than AR3A-JSCC or AR4JA-JSCC for given $p_1=0.04$, we set optimization parameters as follows:
%\begin{enumerate}
%	\item When we want to obtain a better generic protograph than AR3A-JSCC, we set $\bar{p}_1=0.228$ and $(E_s/N_0)^{'}=-5.918$ dB.
%	\item When we want to obtain a better generic protograph than AR4JA-JSCC, we set $\bar{p}_1=0.212$ and $(E_s/N_0)^{'}=-5.767$ dB.
%\end{enumerate}
%
%\begin{equation} \label{AR3A-JSCC}
%	\begin{array}{l}
%		{\textbf{B}}_{\text{AR3A-JSCC}}
%		= \left( {\begin{array}{*{20}{c}}
%				1 & 1 & 2 & 0 & 0 \\
%				0 & 1 & 2 & 1 & 1 \\
%				0 & 2 & 1 & 1 & 1 \\
%		\end{array}} \right) \\ 
%	\end{array}
%\end{equation}
%
%\begin{equation} \label{AR4JA-JSCC}
%	\begin{array}{l}
%		{\textbf{B}}_{\text{AR4JA-JSCC}}
%		= \left( {\begin{array}{*{20}{c}}
%				1 & 0 & 2 & 0 & 0 \\
%				0 & 1 & 3 & 1 & 1 \\
%				0 & 2 & 1 & 2 & 1 \\
%		\end{array}} \right) \\ 
%	\end{array}
%\end{equation}

\subsection{Thresholds of Protomatrices Found}
\subsubsection{Generic protomatrix of size $3 \times 5$ and $p_1=0.04$}
We assume a generic protomatrix of size $3 \times 5$ and use the proposed FSTCT joint optimization method under $p_1=0.04$ 
to search for protomatrices with better thresholds than the two sets of benchmarks separately. Since the protomatrix size is relatively small, we apply 
 the brute force searching approach.
 With $B_1$ as the benchmark, the proposed FSTCT method finds ${\textbf{B}}_{sp\_{opt1}}$ shown in  \eqref{opt1_0} which is repeated below
\begin{equation} \label{opt1}
	\begin{array}{l}
		{\textbf{B}}_{sp\_{opt1}}
		= \left( {\begin{array}{*{20}{c}}
				1 & 1 & 1 & 1 & 1 \\
				0 & 0 & 2 & 0 & 1 \\
				3 & 2 & 2 & 0 & 0 \\
		\end{array}} \right) \\ 
	\end{array};
\end{equation}
and with $B_2$ as the benchmark, the proposed FSTCT method generates 
\begin{equation}\label{opt2}
	\begin{array}{l}
		{\textbf{B}}_{sp\_{opt2}}
		= \left( {\begin{array}{*{20}{c}}
				3 & 2 & 2 & 0 & 0 \\
				0 & 0 & 3 & 1 & 1 \\
				0 & 1 & 1 & 0 & 1 \\
		\end{array}} \right) \\ 
	\end{array}
\end{equation}
\begin{equation}\label{opt3}
	\begin{array}{l}
		{\textbf{B}}_{sp\_{opt3}}
		= \left( {\begin{array}{*{20}{c}}
				1 & 0 & 2 & 0 & 0 \\
				0 & 2 & 3 & 2 & 0 \\
				0 & 1 & 1 & 0 & 3 \\
		\end{array}} \right) \\ 
	\end{array}.
\end{equation}
The source and channel thresholds of these three codes and AR3A-JSCC and AR4JA-JSCC codes are shown in Table~\ref{opt_tabl_sth} and Table~\ref{opt_tabl_cth}, respectively. We can see that these three generic protographs have higher source thresholds than AR3A-JSCC and AR4JA-JSCC codes. 
(Note that Fig.~\ref{ssp_chart} plots the SGP-EXIT chart of $\textbf{B}_{sp\_{opt1}}$ under different $p_1$ values.)
In addition, ${\textbf{B}}_{sp\_{opt1}}$ has the lowest channel threshold ($(E_s/N_0)^{*}$=$-6.1021$ dB) among these five codes, which is about $0.18$ dB lower than the decoding threshold of the AR3A-JSCC code. ${\textbf{B}}_{sp\_{opt2}}$ and ${\textbf{B}}_{sp\_{opt3}}$ have lower channel thresholds compared with the AR4JA-JSCC code.

%
%\begin{table*}[!ht]
%	\caption{The channel thresholds of different protomatrices for given $p_1=0.04$. The shannon limit is $-7.00$ dB}	
%	\centering
%	\begin{tabular}{|c|c|c|c|c|c|c|c|c|}
%		\hline
%		~ & AR3A-JSCC & AR4JA-JSCC & $\textbf{B}_{sp\_{opt1}}$ & $\textbf{B}_{sp\_{opt2}}$ & $\textbf{B}_{sp\_{opt3}}$ & $\textbf{B}_{sp\_{opt1}}^{47}$ & $\textbf{B}_{sp\_{opt2}}^{47}$\\
%		\hline
%		$(E_s/N_0)^{*}$ (dB) & $-5.918$ & $-5.767$ & $-6.102$ & $-5.810$ & $-5.782$ & $-5.909$ & $-6.163$ \\
%		\hline		
%	\end{tabular}
%	\label{opt_tabl_cth}
%\end{table*}

%
%
%Fig.~\ref{ssp_chart} plots the SGP-EXIT chart of $\textbf{B}_{sp\_{opt1}}$ for different $p_1$. The larger the gap between the outer-code curve and the inner-code curve, the faster the decoder converges. The inner-code curve should be above the outer-code curve. The source cannot be reconstructed correctly when these two curves cross. We can see from Fig.~\ref{ssp_chart} that the gap becomes smaller as $p_1$ increases. When $p_1=0.25$, these two curves are closest. This matches the source threshold of $\textbf{B}_{sp\_{opt1}}$ ($p_{1\_{th}}=0.25$) calculated by the SGP-EXIT algorithm.

\begin{figure*}[t]	
	\centering	
	\subfloat[]
	{
		\label{TMDR_B_OPT1}
		\includegraphics[width=3.0in]{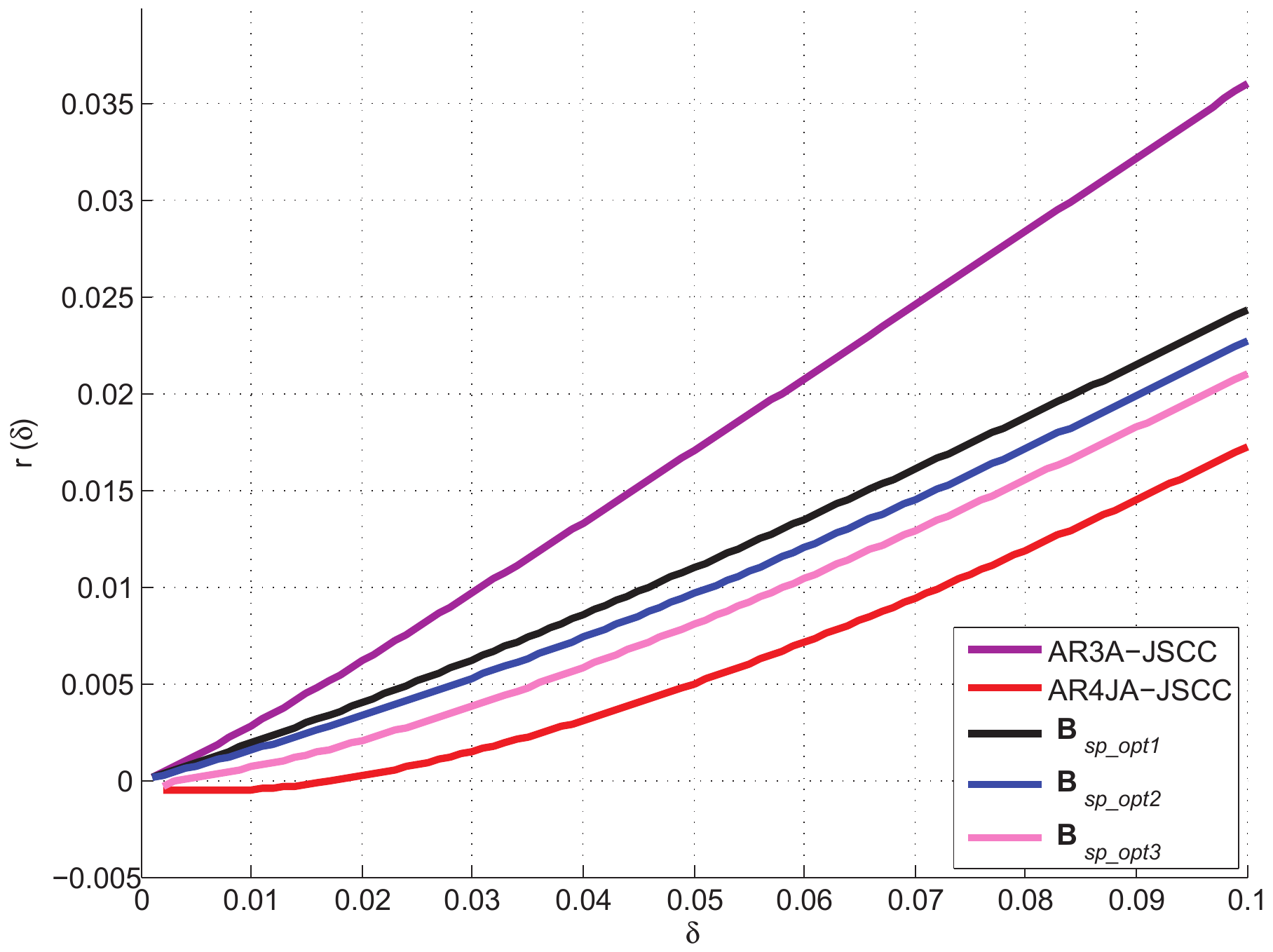}
		\centering
	}
	\subfloat[]
	{
		\label{TMDR_B_OPT2}
		\includegraphics[width=3.0in]{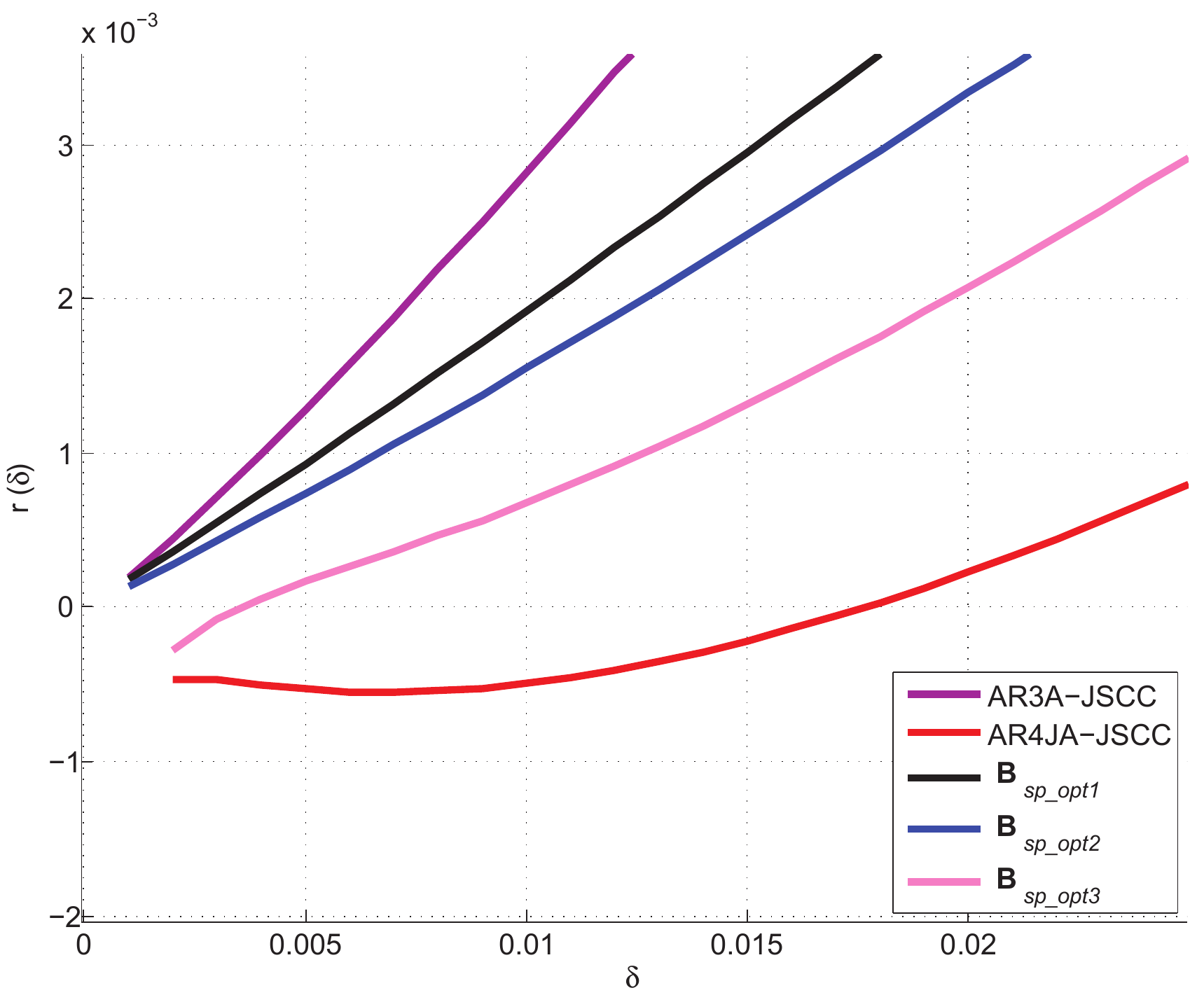}
		\centering
	}	
	\caption{The asymptotic weight distribution curves of different protomatrices. (a) $\delta\in[0,0.1]$; (b) $\delta\in[0,0.025]$.}	
	\label{TMDR_B_OPT}	
\end{figure*}

Using the AWD tool, we plot the asymptotic weight distribution curves of AR3A-JSCC code, AR4JA-JSCC code, ${\textbf{B}}_{sp\_{opt1}}$, ${\textbf{B}}_{sp\_{opt2}}$, 
and ${\textbf{B}}_{sp\_{opt3}}$,  in Fig.~\ref{TMDR_B_OPT}. 
As seen in the figure, AR4JA-JSCC code and ${\textbf{B}}_{sp\_{opt3}}$ have TMDRs while AR3A-JSCC code, ${\textbf{B}}_{sp\_{opt1}}$ and ${\textbf{B}}_{sp\_{opt2}}$ do not have TMDRs. 
The TMDR values of AR4JA-JSCC code and ${\textbf{B}}_{sp\_{opt3}}$ are around $0.017$ and $0.003$, respectively. A larger TMDR value implies a lower error-floor. When a P-LDPC code does not possess a TMDR value, its error-floor performance is hard to be predicted and can only be found out by simulations.	

\subsubsection{Generic protomatrix of size $4 \times 7$ and $p_1=0.04$}
Next, we increase the generic protomatrix size to $4 \times 7$ and use the FSTCT method to search for protomatrices under Benchmarks $B_1$ and $B_2$ separately. Since the protomatrix size is not small, we apply the DE searching approach. 
The parameters $G$, $S$ and $p_c$ in the DE algorithm are set to $800$, $800$ and $0.88$, respectively. 
With Benchmarks {\color{black}$B_1$ and $B_2$}, the FSTCT method finds
protomatrices shown in 
 \eqref{opt1_47} and \eqref{opt2_47}, respectively. 
 {\color{black}
 \begin{equation}\label{opt1_47}
 	\begin{array}{l}
 		{\textbf{B}}_{sp\_{opt1}}^{47}
 		= \left( {\begin{array}{*{20}{c}}
 				1 & 0 & 0 & 1 & 2 & 0 & 1\\
 				0 & 1 & 1 & 1 & 2 & 1 & 0\\
 				0 & 2 & 1 & 3 & 0 & 2 & 0\\
 				1 & 0 & 0 & 2 & 0 & 0 & 0\\
 		\end{array}} \right) \\ 
 	\end{array}
 \end{equation}
\begin{equation}\label{opt2_47}
	\begin{array}{l}
		{\textbf{B}}_{sp\_{opt2}}^{47}
		= \left( {\begin{array}{*{20}{c}}
				0 & 0 & 0 & 2 & 2 & 0 & 1\\
				1 & 0 & 0 & 2 & 0 & 0 & 0\\
				1 & 2 & 0 & 3 & 1 & 1 & 0\\
				0 & 1 & 3 & 1 & 1 & 2 & 0\\
		\end{array}} \right) \\ 
	\end{array}
\end{equation}}
The source and channel thresholds of
${\textbf{B}}_{sp\_{opt1}}^{47}$ and ${\textbf{B}}_{sp\_{opt2}}^{47}$ are shown in Table~\ref{opt_tabl_sth} and Table~\ref{opt_tabl_cth}, respectively. 
Both source thresholds are higher than those of the AR3A-JSCC and AR4JA-JSCC codes. Moreover, $\textbf{B}_{sp\_{opt1}}^{47}$ has a lower channel threshold than {\color{black}AR3A-JSCC} while $\textbf{B}_{sp\_{opt2}}^{47}$ has a lower channel threshold than {\color{black}AR4JA-JSCC}. 
Using the AWD tool, we further find that {\color{black}$\textbf{B}_{sp\_{opt1}}^{47}$ has no TMDR and $\textbf{B}_{sp\_{opt2}}^{47}$ has a TMDR of around $0.007$.}

\begin{table*}[!ht]
	\caption{Channel thresholds of the generic protomatrices at different $p_1$ values. The best channel threshold at each $p_1$ is in bold font while the worst one is in blue color.}	
	\centering
	\begin{tabular}{|c|c||c|c||c|c|c||c|}
		\hline
%		\multirow{2}{*}{$p_1$} & \multirow{2}{*}{Shannon limit} & \multicolumn{6}{c|}{The channel threshold $(E_s/N_0)^{*}$ (dB)} \\
%		\cline{3-8}
  	$p_1$    & Shannon limit & AR3A-JSCC & AR4JA-JSCC & $\textbf{B}_{sp\_{opt1}}$ & $\textbf{B}_{sp\_{opt2}}$ & $\textbf{B}_{sp\_{opt3}}$ & $\textbf{B}_{sp\_{opt4}}$ \\
		\hline
		$0.04$ & $-7.00$ dB & $-5.918$ & $-5.767$ & $\bm{-6.102}$ & $-5.810$ & $-5.782$ & \color{blue}$-4.459$\\
		\hline
		$0.08$ & $-4.19$ dB & $\bm{-3.414}$ & $-3.188$ & $-3.171$ & $-3.027$ & $-3.198$ & \color{blue}$-2.647$\\
		\hline
		$0.12$ & $-2.44$ dB & $\bm{-1.680}$ & $-1.409$ & \color{blue}$-1.151$ & $-1.228$ & $-1.366$ & $-1.239$ \\
		\hline
		$0.16$ & $-1.13$ dB & $-0.094$ & $0.522$ & \color{blue}$0.57$ & $0.193$ & $0.227$ & $\bm{-0.185}$ \\
		\hline
		$0.20$ & $-0.00$ dB & $2.073$ & \color{blue}$3.553$ & $2.30$ & $1.381$ & $2.055$ & $\bm{0.816}$ \\
		\hline		
	\end{tabular}
	\label{opt_tabl_cth_diff}
\end{table*}

\subsubsection{Generic protomatrix of size $3 \times 5$ and $p_1=0.16$}
We design a generic protograph with a size of $3\times 5$ for a relatively large $p_1$, i.e. $p_1=0.16$.  
We use the source threshold and channel threshold of AR3A-JSCC at $p_1=0.16$ as the benchmark. Referring to Tables~\ref{opt_tabl_sth} and~\ref{opt_tabl_cth_diff}, we set
Benchmark $B_3$: $(\bar{p}_1=0.228$, $(E_s/N_0)'$$= -0.094~\text{dB})$.
By using the FSTCT joint optimization method 
 and the brute force searching approach, we obtain 
\begin{equation}
	\begin{array}{l}
		{\textbf{B}}_{sp\_{opt4}}
		= \left( {\begin{array}{*{5}{c}}
				0 & 1 & 1 & 1 & 2 \\
				1 & 1 & 1 & 0 & 1 \\
				0 & 1 & 2 & 2 & 0 \\
		\end{array}} \right) \\ 
	\end{array}.
\end{equation}
with a source threshold of $0.324$ and a channel threshold of $-0.185$~dB at $p_1=0.16$. 
%
%which is within $1$ dB of the Shannon limit. 
%Moreover, 
However, ${\textbf{B}}_{sp\_{opt4}}$ has no TMDR.

\subsubsection{Comparison of channel thresholds at different $p_1$ values}
For the six $3 \times 5$ protomatrices discussed above, namely 
AR3A-JSCC, AR4JA-JSCC,  and ${\textbf{B}}_{sp\_{opt1}}$ to ${\textbf{B}}_{sp\_{opt4}}$, we derive 
their channel thresholds at different $p_1$ values using the PEXIT-JSCC algorithm
and list them in Table~\ref{opt_tabl_cth_diff}. 
We can see that among all codes,
${\textbf{B}}_{sp\_{opt1}}$ has the lowest channel threshold ($-6.102$~dB)  when $p_1=0.04$; 
 AR3A-JSCC has the lowest channel thresholds ($-3.414$ dB and $-1.680$ dB, respectively) 
  when $p_1=0.08$ and $0.12$;
   ${\textbf{B}}_{sp\_{opt4}}$ has the lowest channel thresholds ($-0.185$ dB and $0.816$ dB, respectively) when $p_1=0.16$ and $0.20$. 
   As shown in Table~\ref{opt_tabl_cth_diff}, these lowest channel thresholds
   are within $1$ dB from the Shannon limits.
Moreover, the six $3 \times 5$ protomatrices have different decoding-threshold rankings at different $p_1$ values. The results indicate that a generic protograph optimized at a given $p_1$ does not guarantee the best performance at other $p_1$ values.

{\color{black}\subsubsection{Comparison of thresholds with those of DP-JSCCs}
Table~\ref{dp_th} shows the source thresholds and channel thresholds of 
the optimized double protographs in \cite{liu2020joint} (denoted by
$\textbf{B}_J^{opt\_1}$ to $\textbf{B}_J^{opt\_4}$) when $p_1=0.04$.
Comparing the results with those in Table~\ref{opt_tabl_cth}
and Table~\ref{opt_tabl_sth} indicate that our constructed P-JSSCs 
can achieve better thresholds than $\textbf{B}_J^{opt\_1}$ to $\textbf{B}_J^{opt\_4}$. 
}

\begin{table}[!t]
	\caption{{\color{black} The source thresholds and channel thresholds of 
	the optimized double protographs in \cite{liu2020joint} when $p_1=0.04$. 
	{\color{black}The size of each double-protograph is $5 \times 9$.} The Shannon limit is $-7.00$ {\upshape d}B}}	
	\centering
	\begin{tabular}{|c|c|c|c|c|}
		\hline
		~ & $\textbf{B}_J^{opt\_1}$ & $\textbf{B}_J^{opt\_2}$ & $\textbf{B}_J^{opt\_3}$ & $\textbf{B}_J^{opt\_4}$\\
		\hline
		$p_{1\_{th}}$ & $0.082$ & $0.082$ & $0.144$ & $0.136$ \\
	    \hline
		$(E_s/N_0)^{*}$ (dB) & $-5.130$ & $-5.398$ & $-5.267$ & $-5.571$ \\
		\hline 	
	\end{tabular}
	\label{dp_th}
\end{table}

\subsection{Error Performance}
In addition to theoretical analysis, computer simulations are performed. 
The encoding and decoding processes have been described in 
Sect.~\ref{sect:P-JSCC encoder} and Sect. \ref{sect:P-JSCC decoder}, respectively.
 We denote the number of source symbols in a frame by $N_s=n_rz$
and set the maximum number of decoding iterations 
 to $I_{\max}=200$.
Three types of error rates are recorded. 
\begin{itemize}
\item Source symbol error rate (SSER) is evaluated by 
 comparing the original  source symbols with the recovered  source symbols.
\item Transmitted bit error rate (TBER) is evaluated by 
 comparing the code bits sent through the channel
 with the corresponding recovered bits. 
 \item Frame error rate (FER) is evaluated by 
 comparing the original JSCC codeword 
 with the recovered codeword. 
\end{itemize}
%The first one is the  The second one is the transmitted bit error rate (TBER) which is obtained by comparing the transmitted bits with the reconstructed bits corresponding to them at the receiver. The third one is the frame error rate (FER). 
The error rate results are recorded if 
(i) the number of  frames simulated exceeds $10^{5}$, or (ii) the number of error frames exceeds $50$ and the number of  frames simulated is no smaller than $5000$. 

%When $N_s=12800$ and $E_s/N_0=-5.1$ dB, only $6$ frame errors are collected for $\textbf{B}_{sp\_{opt3}}$ and the total number of simulated frames is 498145.

\begin{figure*}[!ht]	
	\centering	
	\subfloat[]
	{
		\label{res1_0p04_1}
		\includegraphics[width=3.5in]{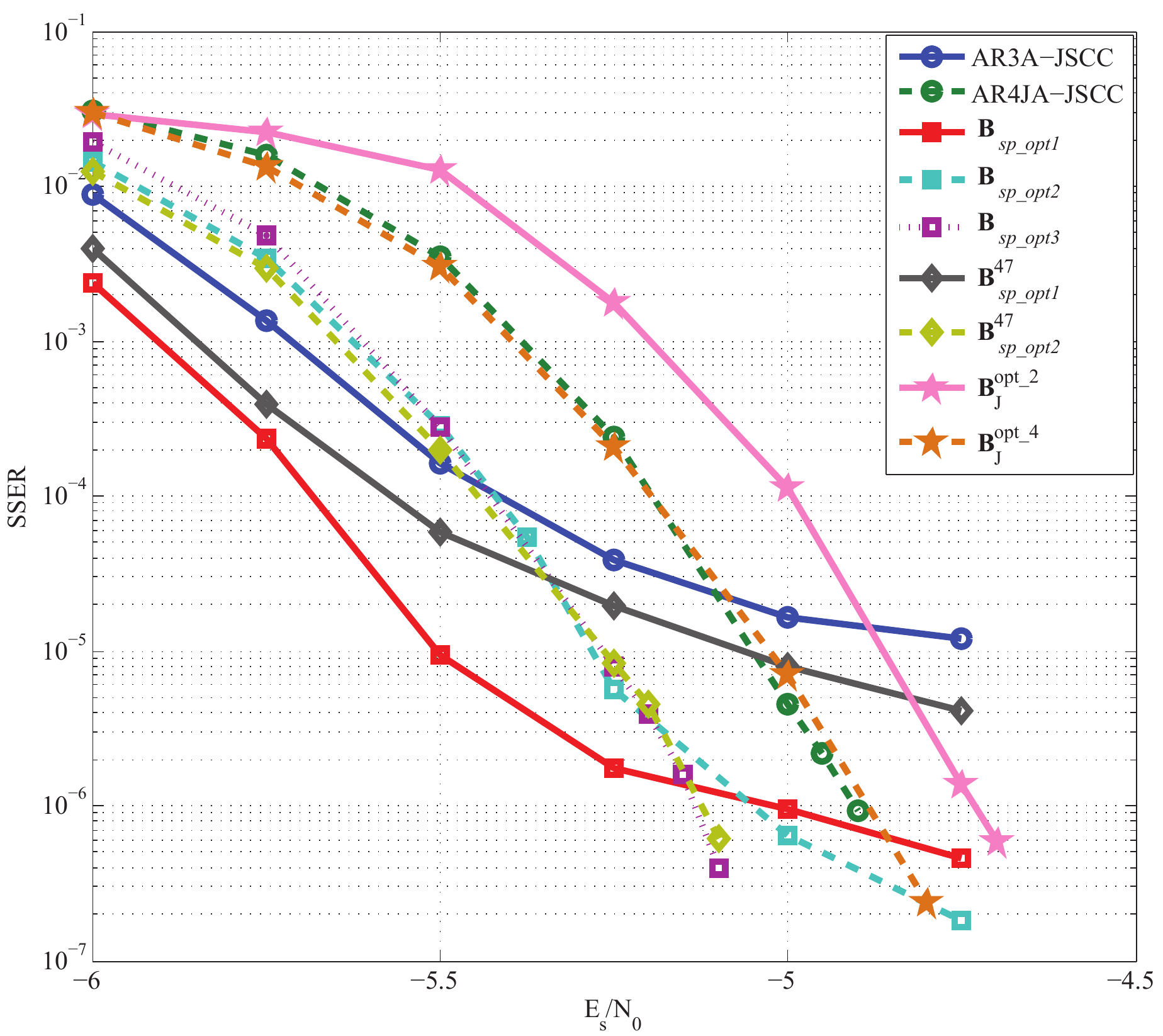}
		\centering
	}
	\subfloat[]
	{   
		\label{res1_0p04_2}
		\includegraphics[width=3.5in]{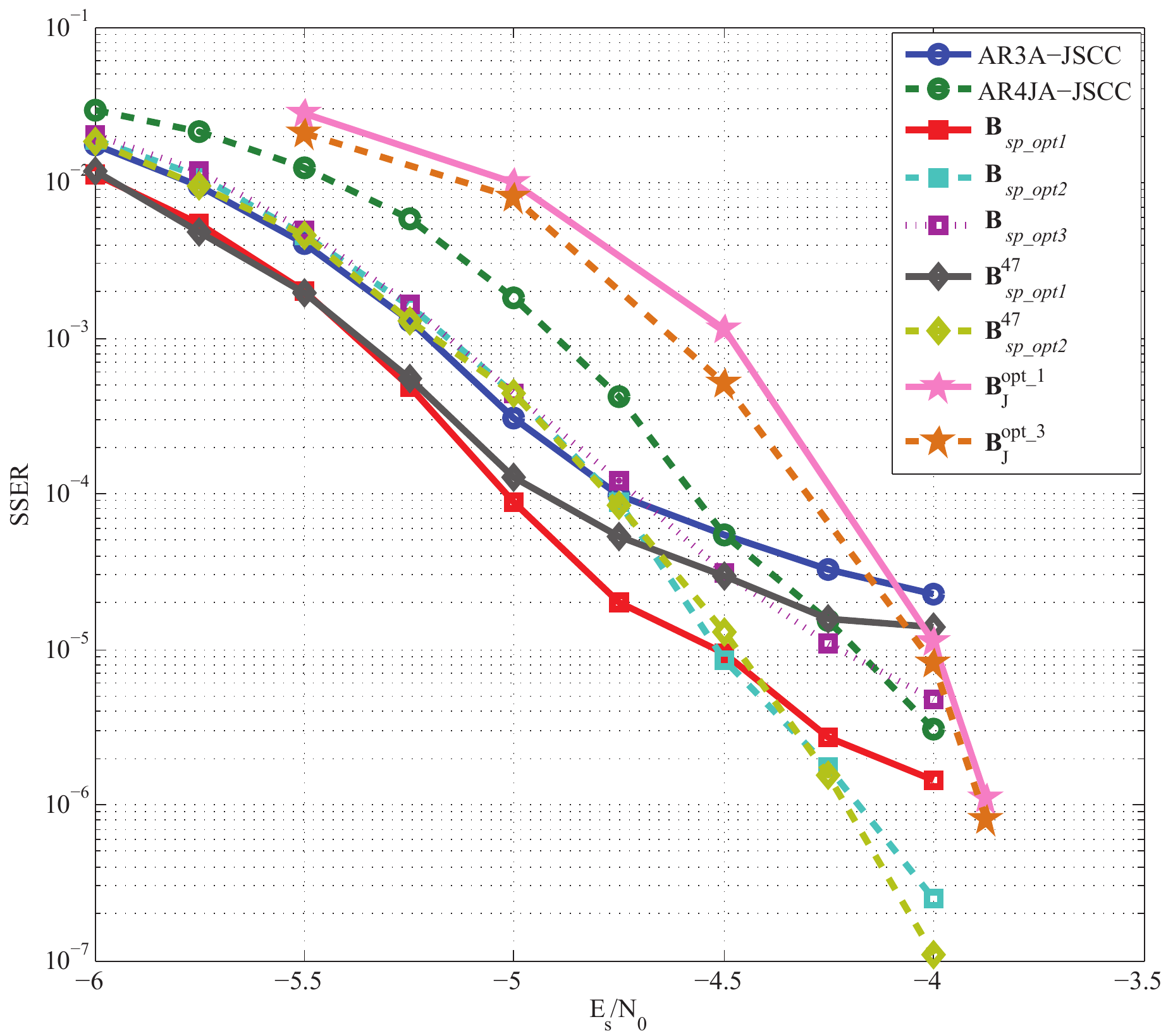}
		\centering
	}	
	\caption{SSER performance comparison at $p_1=0.04$.
	(a) 
	 $N_s=12800$ (i.e. $z=6400$) for all generic $3\times 5$ protographs, and $N_s=12864$ (i.e. $z=4288$) for all generic $4\times 7$ protographs.
Starting from $E_s/N_0=-5.0$ dB, $-5.25$ dB, $-5.0$ dB and $-5.2$ dB, respectively, the numbers of error frames of AR4JA-JSCC, $\textbf{B}_{sp\_{opt3}}$, $\textbf{B}_{sp\_{opt1}}^{47}$ and $\textbf{B}_{sp\_{opt2}}^{47}$ recorded are below $50$. 
(b) $N_s=3200$ (i.e. $z=1600$)  for all generic $3\times 5$ protographs, and $N_s=3264$ (i.e. $z=1088$)  for all generic $4\times 7$ protographs.
Starting from $E_s/N_0=-4.25$ dB, $-4.25$ dB, $-4.5$ dB and $-4.25$ dB, respectively,  
the numbers of error frames of AR4JA-JSCC, $\textbf{B}_{sp\_{opt2}}$, $\textbf{B}_{sp\_{opt3}}$ and $\textbf{B}_{sp\_{opt2}}^{47}$ recorded are below $50$.}	
	\label{res1_0p04}	
\end{figure*}

%Fig.~\ref{res1_0p04}\subref{res1_0p04_1} and Fig.~\ref{res1_0p04}\subref{res1_0p04_2} 
Fig.~\ref{res1_0p04} plots the SSER performance of 
all our constructed protographs optimized at $p_1=0.04$.
We also plot the results of AR3A-JSCC, AR4JA-JSCC and 
the optimized double protographs in \cite{liu2020joint} (i.e.,  
$\textbf{B}_J^{opt\_1}$ to $\textbf{B}_J^{opt\_4}$) for comparison. 
From Fig.~\ref{res1_0p04}\subref{res1_0p04_1}  where $N_s$ is around $12800$,
we can observe that 
\begin{enumerate}
\item $\textbf{B}_{sp\_{opt1}}$ and $\textbf{B}_{sp\_{opt1}}^{47}$ outperform
AR3A-JSCC in both the 
waterfall and high-SNR regions while all three codes have error floors;
\item $\textbf{B}_{sp\_{opt2}}$, $\textbf{B}_{sp\_{opt3}}$ and $\textbf{B}_{sp\_{opt2}}^{47}$ have, respectively, 
 $0.20$~dB, $0.25$~dB and $0.25$~dB gains 
 over AR4JA-JSCC at a BER of $10^{-6}$
\item  $\textbf{B}_{sp\_{opt3}}$ and $\textbf{B}_{sp\_{opt2}}^{47}$ has no error floor 
down to a BER of $10^{-6}$;  
\item optimized double protographs $\textbf{B}_J^{opt\_2}$ and $\textbf{B}_J^{opt\_4}$ in \cite{liu2020joint} are outperformed by all the generic protographs in the waterfall region and they do not have error floors 
down to a BER of $10^{-6}$;
\item error floors exist or start to emerge for AR3A-JSCC, $\textbf{B}_{sp\_{opt1}}$, $\textbf{B}_{sp\_{opt2}}$ and $\textbf{B}_{sp\_{opt1}}^{47}$ which do not have TMDRs. 
\end{enumerate}
When   $N_s$ is reduced to around $3200$, we can observe 
from Fig.~\ref{res1_0p04}\subref{res1_0p04_2} that the performance of these codes are degraded.
Yet, $\textbf{B}_{sp\_{opt1}}$, $\textbf{B}_{sp\_{opt2}}$ and $\textbf{B}_{sp\_{opt2}}^{47}$ still outperform the optimized double protographs $\textbf{B}_J^{opt\_1}$ and $\textbf{B}_J^{opt\_3}$ in \cite{liu2020joint} down to a BER of around $10^{-6}$.

%\begin{table*}[!t]
%	\caption{No. of error frames \& total no. of frames simulated for different codes and specific $E_s/N_0$}	
%	\centering
%	\begin{tabular}{|c|c|c|c|c|}
%		\hline
%		$E_s/N_0$ & $p_1$ & Code & Number of error frames & Total number of frames\\
%		\hline
%		$-2.45$ (dB) & $0.08$ & AR4JA-JSCC & $5$ & $586274$\\
%		\hline	
%		$-2.4$ (dB) & $0.08$ & $\textbf{B}_{sp\_{opt1}}$ & $28$ & $640295$\\
%		\hline
%		$-2.35$ (dB) & $0.08$ & $\textbf{B}_{sp\_{opt2}}$ & $4$ & $763127$\\
%		\hline	
%		$-2.5$ (dB) & $0.08$ & $\textbf{B}_{sp\_{opt3}}$ & $24$ & $683218$\\
%		\hline	
%		$-0.9$ (dB) & $0.12$ & AR3A-JSCC & $25$ & $563320$\\
%		\hline	
%		$-0.75$ (dB) & $0.12$ & AR4JA-JSCC & $25$ & $748348$\\
%		\hline	
%		$-0.5$ (dB) & $0.12$ & $\textbf{B}_{sp\_{opt1}}$ & $22$ & $494043$\\
%		\hline	
%		$-0.52$ (dB) & $0.12$ & $\textbf{B}_{sp\_{opt2}}$ & $3$ & $784960$\\
%		\hline	
%		$0.65$ (dB) & $0.16$ & AR3A-JSCC & $11$ & $610737$\\
%		\hline	
%		$1.30$ (dB) & $0.16$ & AR4JA-JSCC & $25$ & $465042$\\
%		\hline	
%		$0.9$ (dB) & $0.16$ & $\textbf{B}_{sp\_{opt2}}$ & $3$ & $764614$\\
%		\hline			
%	\end{tabular}
%	\label{opt_tab_dif}
%\end{table*}

\begin{figure*}[htbp]
	\centering
	\subfloat[$p_1=0.04$ ]{
		\includegraphics[width=3.4in]{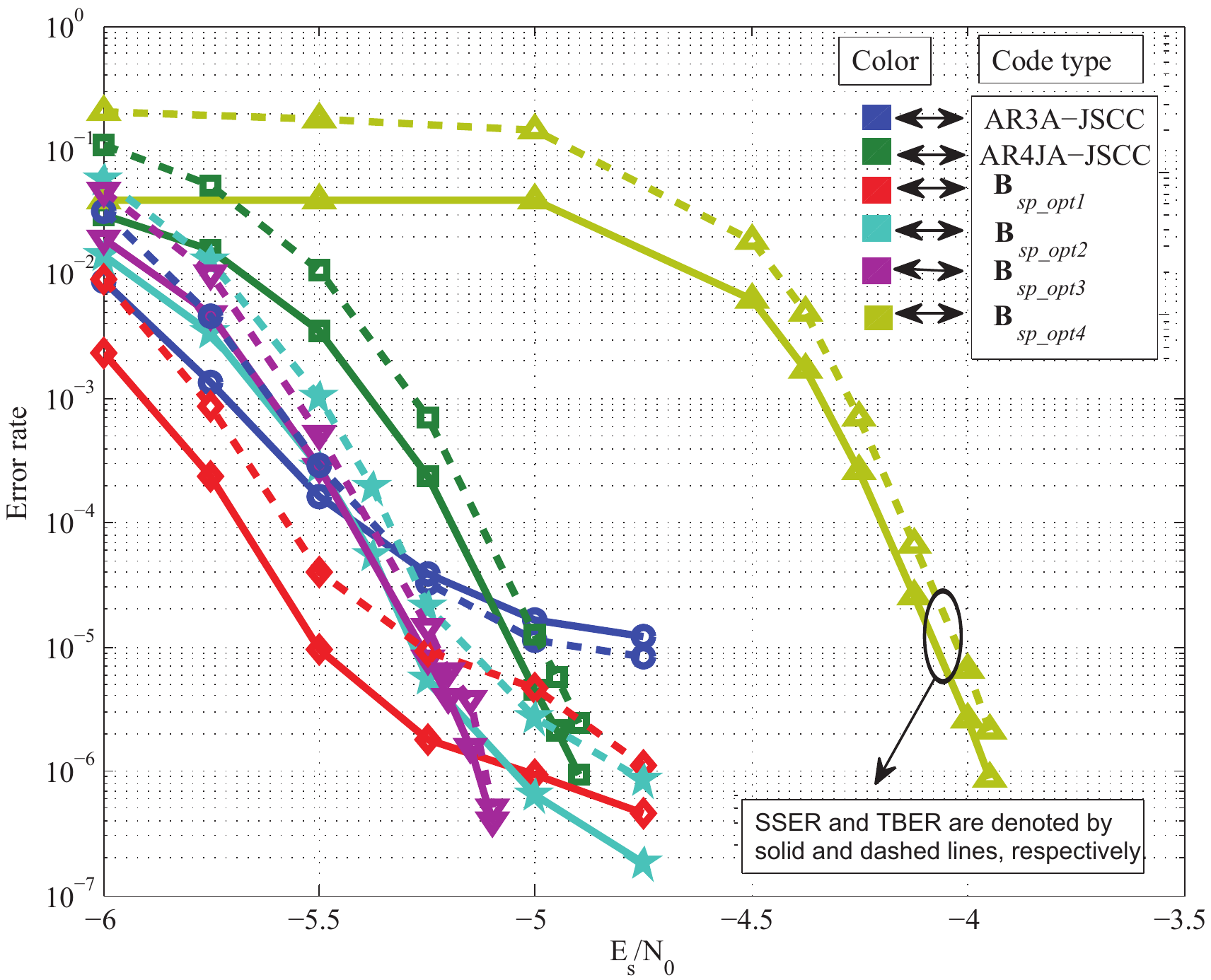}
		\label{BER_dif_p.a}
	}
	\subfloat[$p_1=0.08$ ]{
		\includegraphics[width=3.4in]{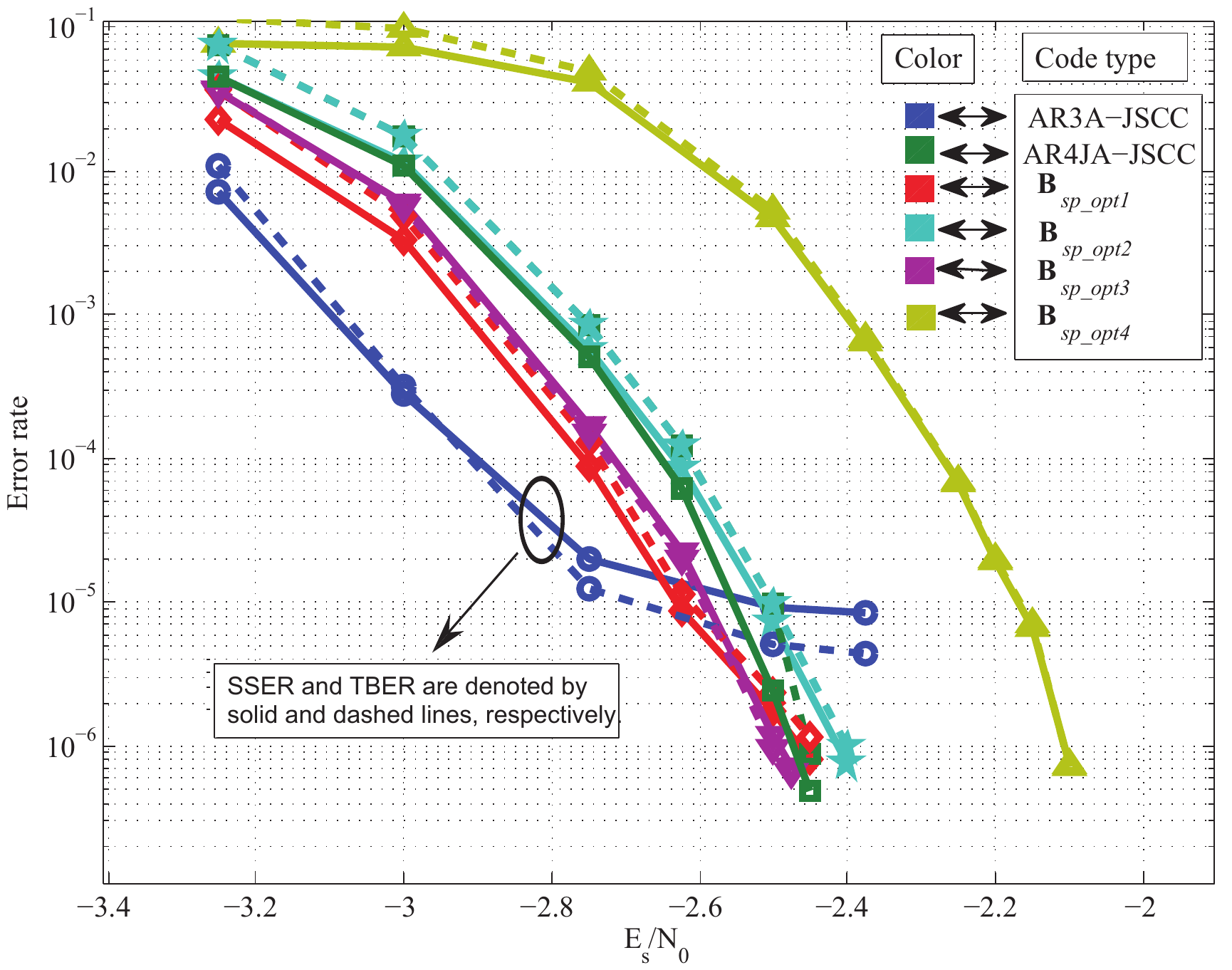} 
		\label{BER_dif_p.b}
	} \vskip -8pt
	\quad   
	\subfloat[$p_1=0.12$ ]{
		\includegraphics[width=3.4in]{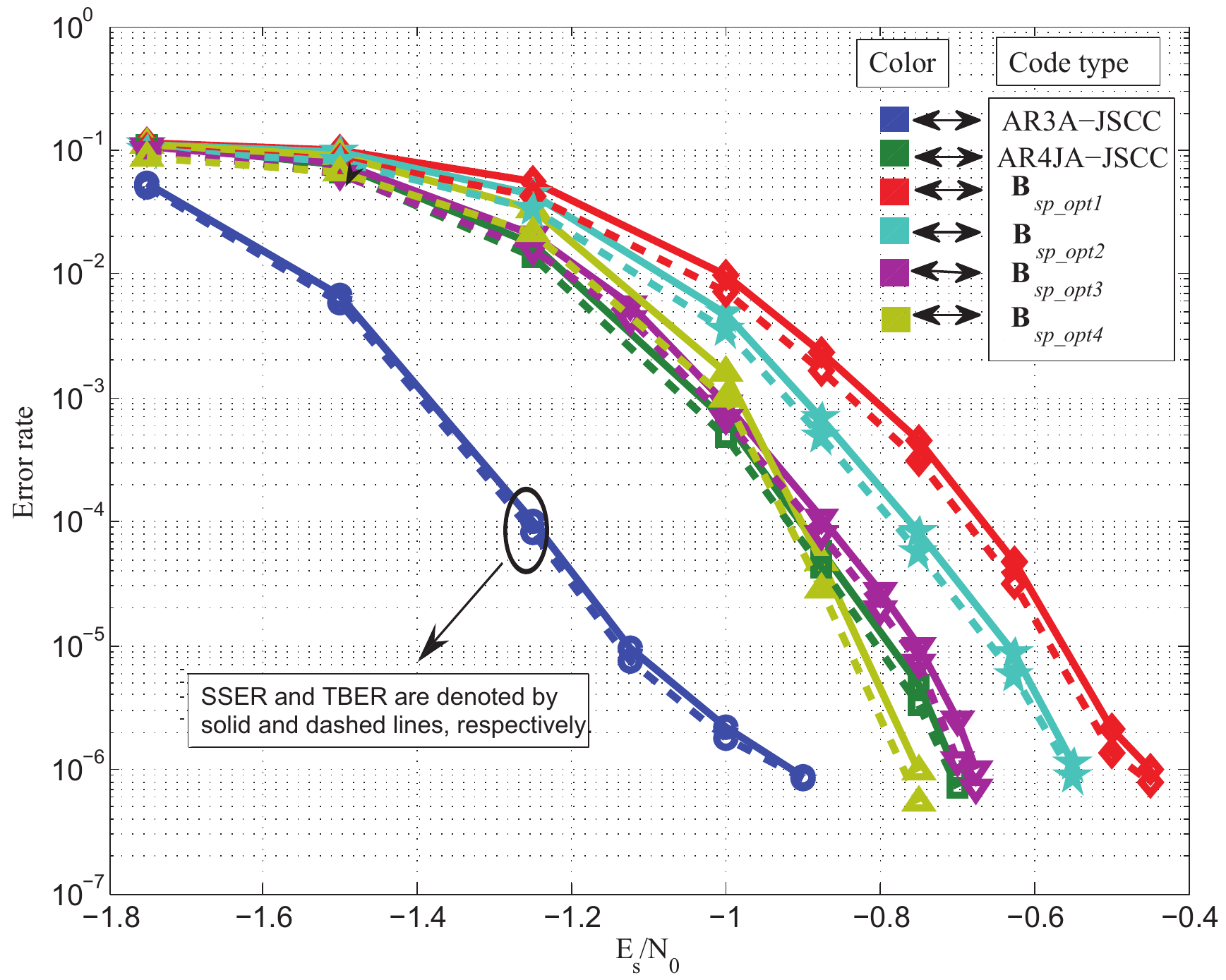}
		\label{BER_dif_p.c}
	}    
	\subfloat[$p_1=0.16$]{
		\includegraphics[width=3.4in]{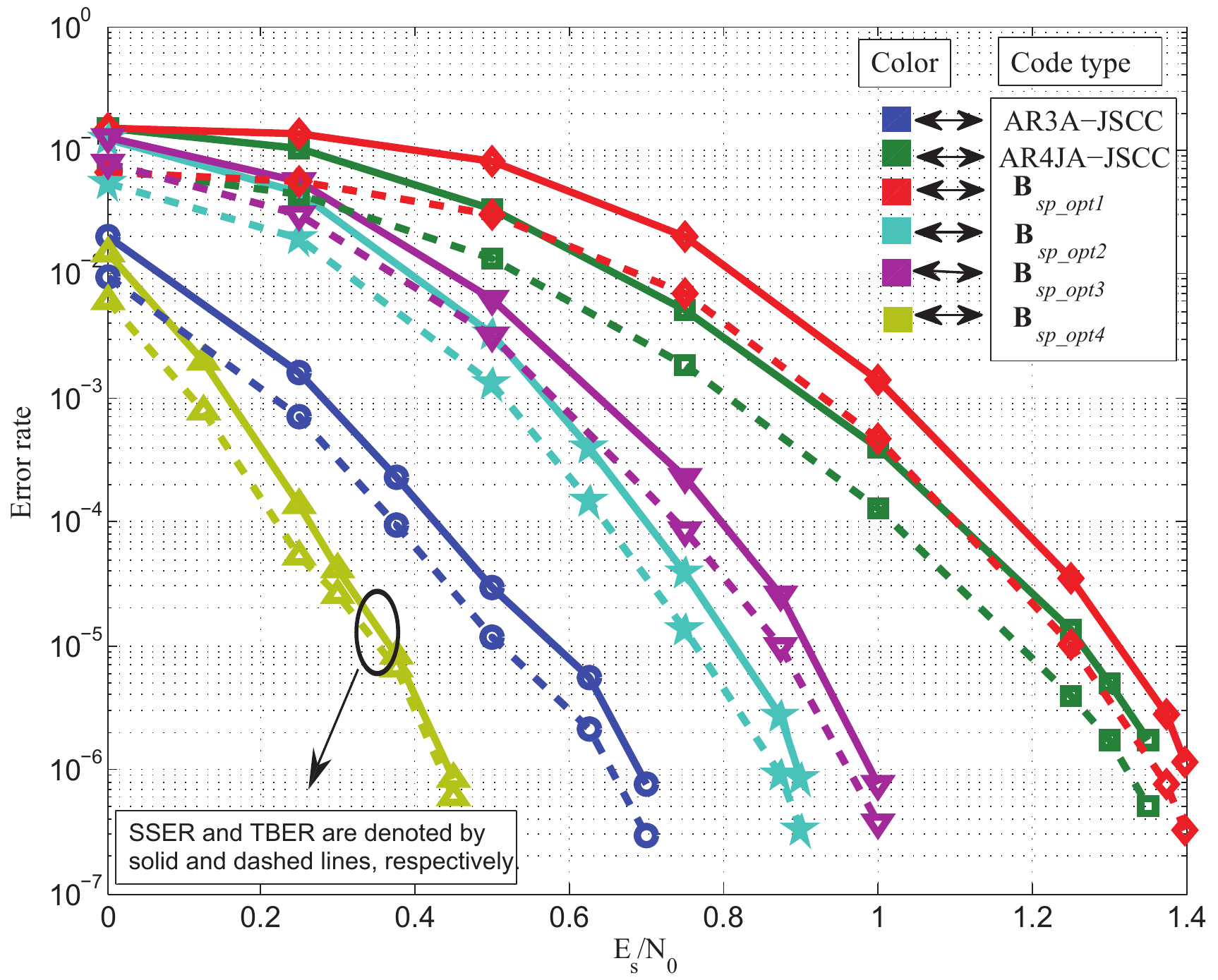}
		\label{BER_dif_p.d}
	}    
    \quad
	\subfloat[$p_1=0.2$]{
		\includegraphics[width=3.4in]{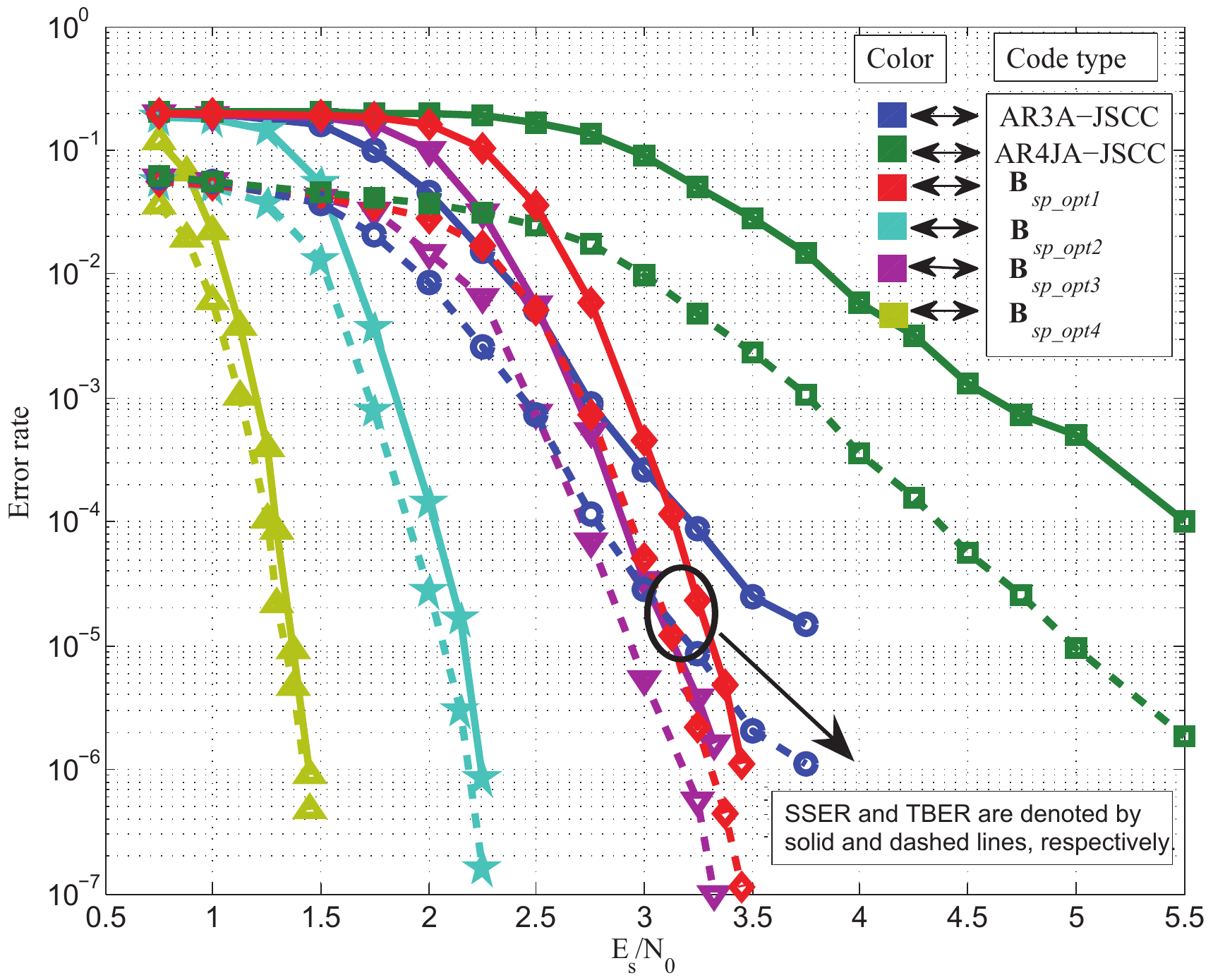}
		\label{BER_dif_p.e}
	} 
    \subfloat[\small \rm The table lists the starting $E_s/N_0$ values, from which 
     the numbers of error frames recorded are below $50$.]{ \includegraphics[width=3.4in]{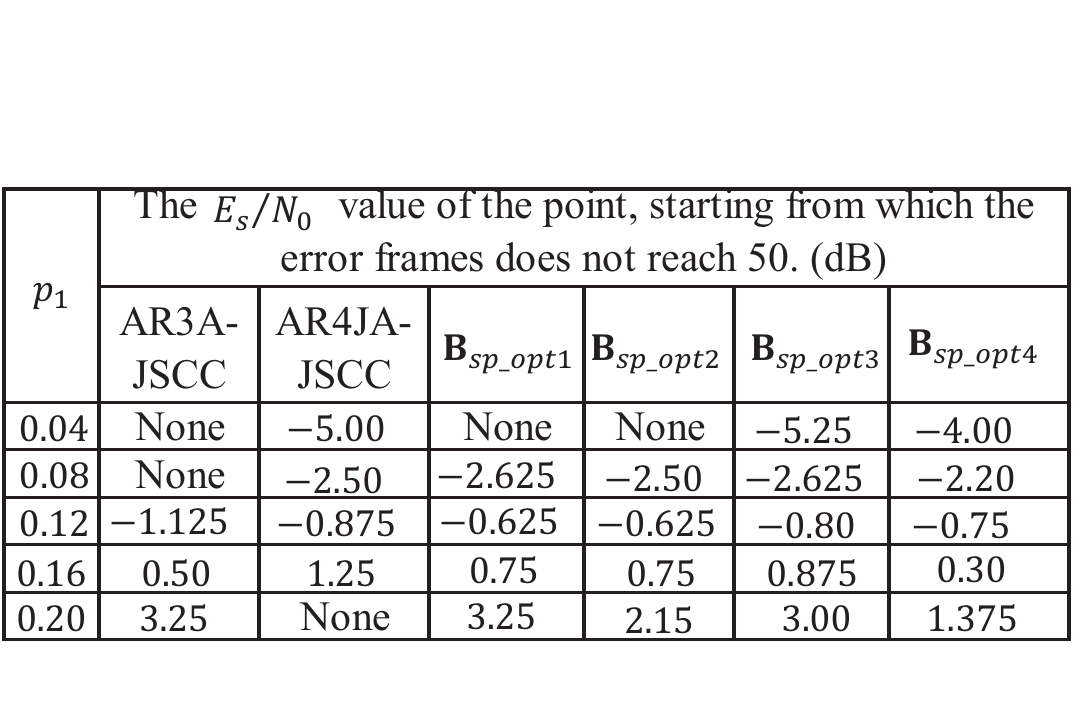}
    	\label{BER_dif_p.f}
    }   
	\caption{Source symbol error rate (SSER denoted by solid lines) and transmitted bit error rate (TBER denoted by dashed lines) performance of AR3A-JSCC, AR4JA-JSCC and the optimized codes constructed  ($\textbf{B}_{sp\_{opt1}}$ to $\textbf{B}_{sp\_{opt4}}$). $N_s=12800$. The source thresholds of AR3A-JSCC, AR4JA-JSCC, $\textbf{B}_{sp\_{opt1}}$, $\textbf{B}_{sp\_{opt2}}$, $\textbf{B}_{sp\_{opt3}}$ and $\textbf{B}_{sp\_{opt4}}$ are $0.228$, $0.212$, $0.25$, $0.275$, $0.242$ and $0.324$, respectively.
	}
	\label{BER_dif_p}
\end{figure*}

%Table~\ref{opt_tab_dif} lists the number of error frames and total no. of simulation frames when less than 50 error frames are collected. 

Fig.~\ref{BER_dif_p} depicts the SSER and TBER performance of $\textbf{B}_{sp\_{opt1}}$, $\textbf{B}_{sp\_{opt2}}$, $\textbf{B}_{sp\_{opt3}}$, $\textbf{B}_{sp\_{opt4}}$, AR3A-JSCC code, AR4JA-JSCC code when $N_s=12800$ and $p_1=0.04, 0.08, 0.12, 0.16$ and $0.20$. 
We can observe that the error performance of all codes  degrades in general as $p_1$ increases from $0.04$ to $0.20$. 
It is because the initial LLR information of the source symbols decreases as $p_1$ increases.
Hence in order to achieve the same error performance,
a larger $E_s/N_0$ is required to compensate the reduction 
in initial information as $p_1$ increases.

Referring to Fig.~\ref{BER_dif_p}(a) where $p_1=0.04$, the initial LLR information of the source symbols is relatively large
compared with the channel LLR information of the transmitted bits.  
The source symbols therefore have a higher chance of being decoded correctly even 
when the transmitted bits are decoded wrongly.  Thus 
SSER is better (lower) than TBER for the same $E_s/N_0$ in the given SNR range.
As $p_1$ increases, the initial LLR information of the source symbols decreases.
The source symbols rely more heavily on the channel LLRs of the transmitted bits 
for correct decoding. When the transmitted bits cannot be decoded correctly
and hence cannot pass reliable information to the source symbols, the source symbols become even less likely to be decoded correctly. Therefore when $p_1=0.20$,
Fig.~\ref{BER_dif_p}(e) shows that SSER is worse (higher) than TBER for the same $E_s/N_0$ in the given SNR range.
Based on the same arguments, we can also conclude the following.
\begin{itemize}
\item When $p_1$ is low  (e.g., $0.04$), 
the source symbols have a higher chance of being decoded correctly
but the lack of TMDRs of the codes causes an error floor
(see Table~\ref{opt_tabl_cth} and the error curves for AR3A-JSCC, $\textbf{B}_{sp\_{opt1}}$
and $\textbf{B}_{sp\_{opt2}}$ in Fig.~\ref{BER_dif_p}(a)).
\item When $p_1$ becomes large (e.g., $0.20$) and approaches the source thresholds
of the codes (e.g., AR3A-JSCC and AR4JA-JSCC have source thresholds of $0.228$ and $0.212$, respectively), 
the low initial LLR information of the source symbols
cannot ensure successful decoding of the symbols
even at high SNR, causing error floors to occur (see the error curves for AR3A-JSCC and AR4JA-JSCC  in Fig.~\ref{BER_dif_p}(e)). 
\end{itemize}
Moreover, at the waterfall region, we can observe
\begin{enumerate}
\item at $p_1=0.04$, $\textbf{B}_{sp\_{opt1}}$ performs the best while $\textbf{B}_{sp\_{opt4}}$ performs the worst;
\item at $p_1=0.08$, AR3A-JSCC performs the best while $\textbf{B}_{sp\_{opt4}}$ performs the worst;
\item at $p_1=0.12$, AR3A-JSCC performs the best while $\textbf{B}_{sp\_{opt1}}$ performs the worst;
\item at $p_1=0.16$, $\textbf{B}_{sp\_{opt4}}$ performs the best while $\textbf{B}_{sp\_{opt1}}$ performs the worst; and
\item at $p_1=0.20$, $\textbf{B}_{sp\_{opt4}}$ performs the best while AR4JA-JSCC performs the worst;
\end{enumerate}
The relative error performances of the codes  therefore
match with the channel thresholds 
listed in Table~\ref{opt_tabl_cth_diff}.

\begin{figure*}[htbp]
	\centering
	\subfloat[$p_1=0.04$]{
		\includegraphics[width=3.4in]{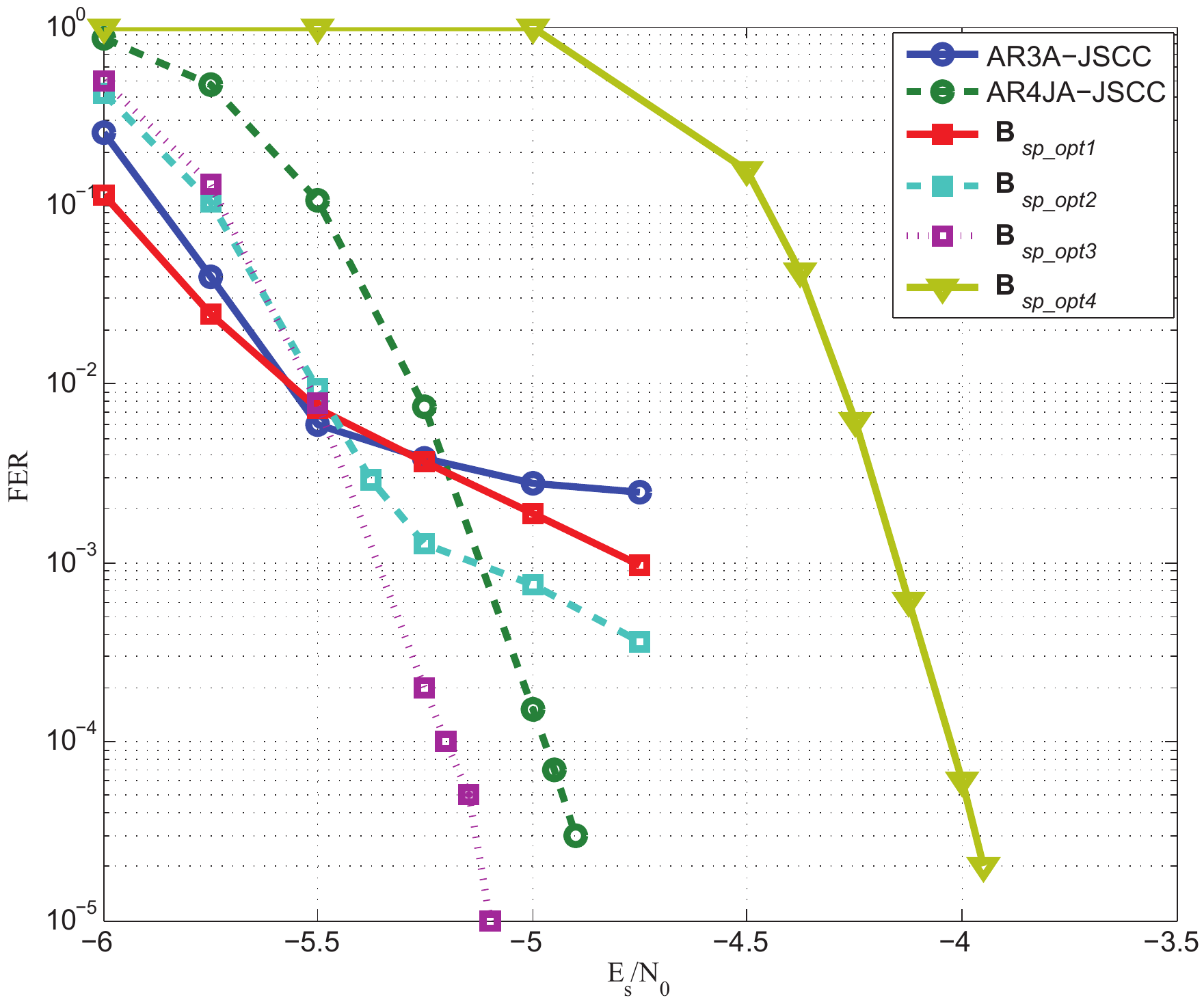}
	} 
    \label{FER_dif_p.a}
	\subfloat[$p_1=0.08$ \label{FER_dif_p.b}]{
		\includegraphics[width=3.2in]{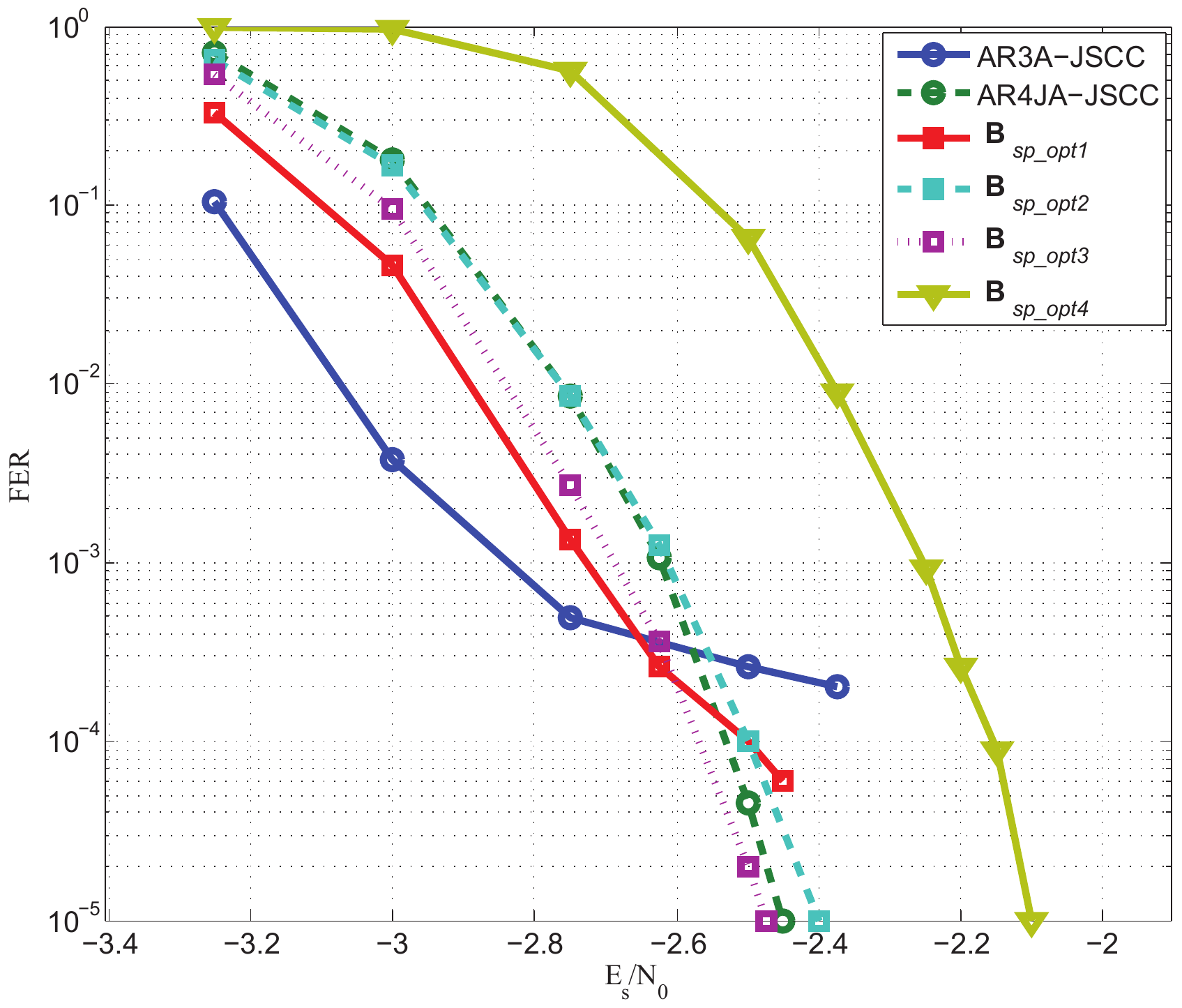}
	}      
   \quad 
	\subfloat[$p_1=0.12$ \label{FER_dif_p.c}]{
		\includegraphics[width=3.4in]{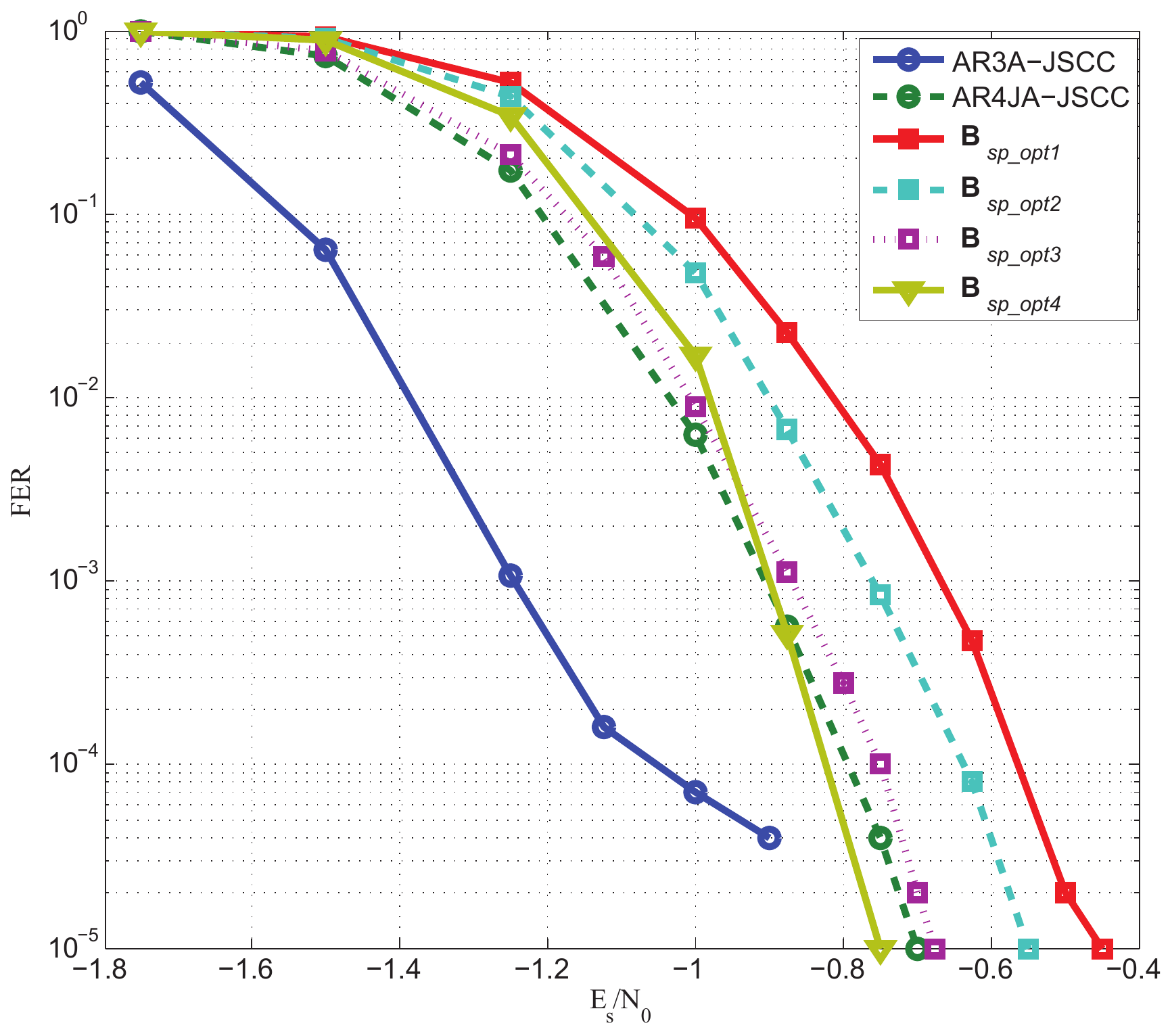}
	}   
	\subfloat[$p_1=0.16$ \label{FER_dif_p.d}]{
		\includegraphics[width=3.4in]{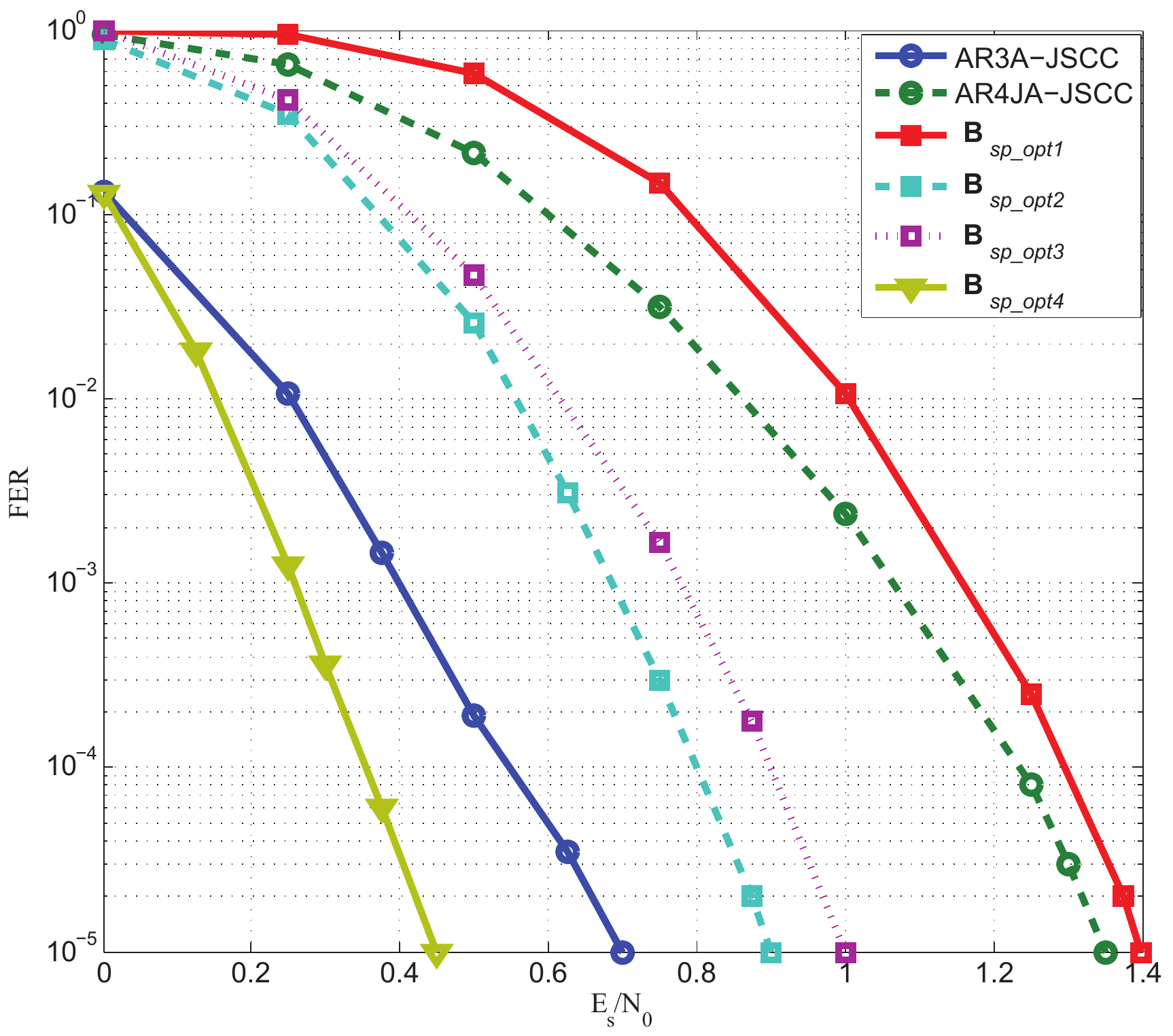}
	}    
    \quad
	\subfloat[$p_1=0.20$ \label{FER_dif_p.e}]{
		\includegraphics[width=3.4in]{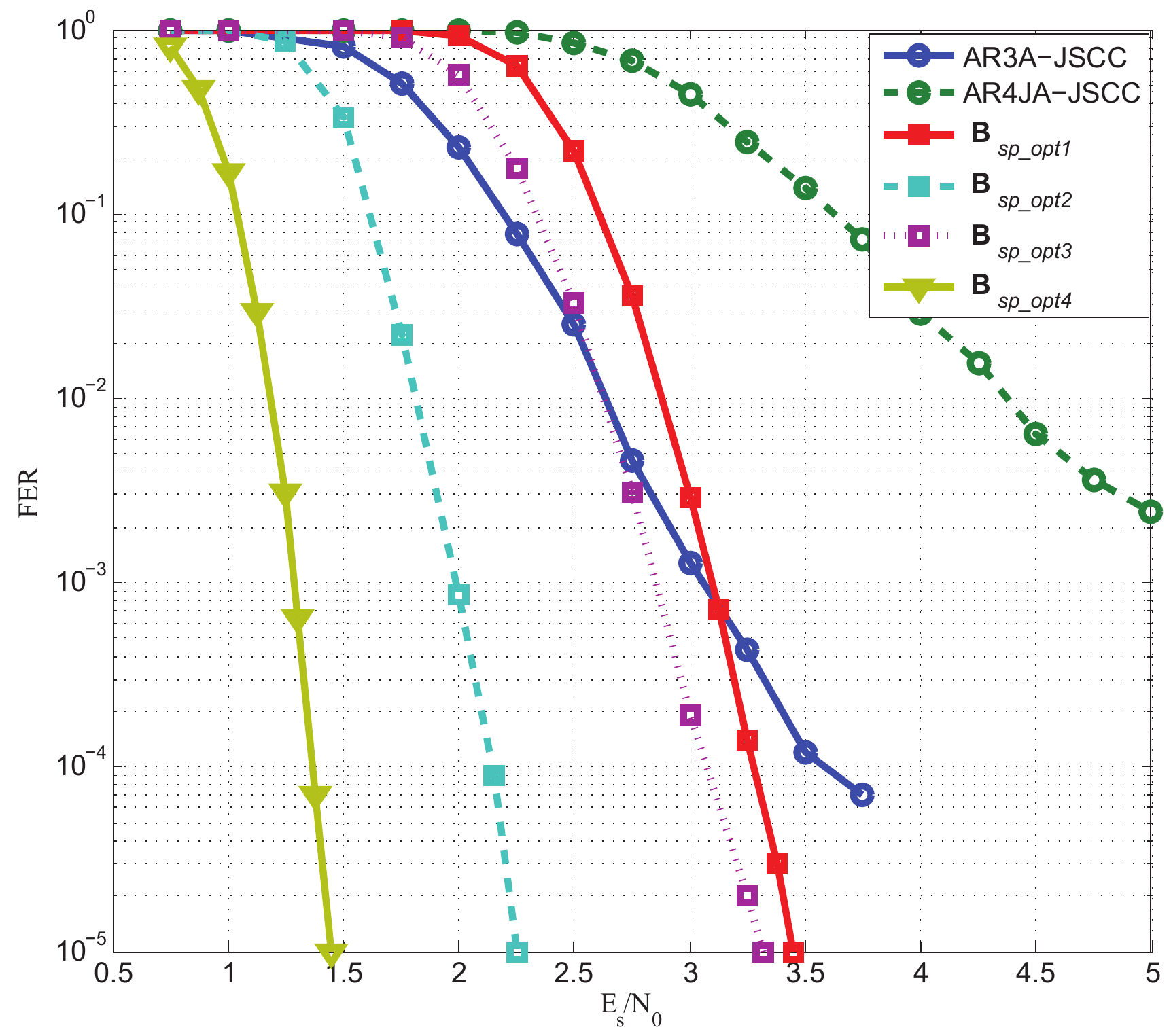}
	}    
	\caption{Frame error rate (FER) performance of AR3A-JSCC, AR4JA-JSCC and the optimized codes constructed ($\textbf{B}_{sp\_{opt1}}$ to $\textbf{B}_{sp\_{opt4}}$). $N_s=12800$. The source thresholds of AR3A-JSCC, AR4JA-JSCC, $\textbf{B}_{sp\_{opt1}}$, $\textbf{B}_{sp\_{opt2}}$, $\textbf{B}_{sp\_{opt3}}$ and $\textbf{B}_{sp\_{opt4}}$ are $0.228$, $0.212$, $0.25$, $0.275$, $0.242$ and $0.324$, respectively.}
	\label{FER_dif_p}
\end{figure*}

Fig.~\ref{FER_dif_p} depicts the frame error rate (FER) performance of 
$\textbf{B}_{sp\_{opt1}}$, $\textbf{B}_{sp\_{opt2}}$, $\textbf{B}_{sp\_{opt3}}$, $\textbf{B}_{sp\_{opt4}}$, AR3A-JSCC code, AR4JA-JSCC code when $N_s=12800$ and $p_1=0.04, 0.08, 0.12, 0.16$ and $0.20$. The characteristics of the curves can be explained similarly using the aforementioned arguments.

\section{Conclusions}\label{sect:conclusion}
In this paper, 
we propose a JSCC system based on a generic protograph,
namely protograph-based JSCC (P-JSCC).
We present a PEXIT-JSCC algorithm to evaluate the
channel threshold
and propose a source generic protograph EXIT (SGP-EXIT) algorithm
to evaluate the source threshold of a P-JSCC.
Based on the PEXIT-JSCC and SGP-EXIT algorithms, 
we further propose a joint optimization method to design 
P-JSCCs (i.e., generic protographs) with good channel and source thresholds.
In terms of theoretical thresholds and error rates, the performance of the P-JSCCs constructed by the 
joint optimization method are found to outperform JSCCs based on 
double protographs. 

We further find that the source symbol error rate (SSER) 
 is better (lower) than the transmitted bit error rate (TBER) 
 when the probability of ``1" in the source sequence $p_1$
 is small, and vice versa. 
Moreover, error floors are caused by
(i)  a lack of 
 TMDRs in the protographs when $p_1$
 is small, and 
 (ii) small initial LLR information of the source symbols when $p_1$ is large and 
 approaches the source thresholds.

\bibliographystyle{IEEEtran}
\bibliography{BibTeX}

\end{document}